\documentclass{nature}

\usepackage{amsmath}
\usepackage{ctable}
\usepackage{multirow}
\usepackage{colortbl}
\usepackage{tabularx}
\usepackage{chapterbib}
\usepackage{float}
\usepackage{caption}
\usepackage{subcaption}
\usepackage{capt-of}
\usepackage{multibib}
\usepackage{pdfpages}

\usepackage{graphicx}
\makeatletter
\let\saved@includegraphics\includegraphics
\AtBeginDocument{\let\includegraphics\saved@includegraphics}
\renewenvironment*{figure}{\@float{figure}}{\end@float}
\makeatother

\graphicspath{{./figures/}}

\newcommand{\acl}{NASCUP}
\newcommand{\eg}{{\it e.g.}}
\newcommand{\ie}{{\it i.e.}}

\newcommand{\ttA}{\texttt{A}}
\newcommand{\ttC}{\texttt{C}}
\newcommand{\ttG}{\texttt{G}}
\newcommand{\ttT}{\texttt{T}}
\newcommand{\ttS}{\texttt{*}}
\newcommand{\bn}{\mathbf{n}}
\newcommand{\bm}{\mathbf{m}}
\newcommand{\bx}{\mathbf{x}}
\newcommand{\by}{\mathbf{y}}
\newcommand{\nts}{\texttt{A},\texttt{C},\texttt{G},\texttt{T}}

\definecolor {shade}{rgb}{0.9,0.9,0.9}
\def\CC{C++}

\renewcommand*\figurename{Fig.}
\newcommand*\supfigurename{Supplementary Fig.}
\newcommand*\suptablename{Supplementary Table}
\newcommand*\supnotename{Supplementary Note}

\title{NASCUP: Nucleic Acid Sequence Classification by \\
Universal Probability}

\author{Sunyoung Kwon$^{1}$, Gyuwan Kim$^{1}$, Byunghan Lee$^{1}$, Jongsik Chun$^{2,3}$, Sungroh Yoon$^{1,2,3*}$, \& Young-Han Kim$^{4*}$}

\begin{document}

\maketitle

\begin{affiliations}
\item Department of Electrical and Computer Engineering, Seoul National University, Seoul, Korea.
\item School of Biological Sciences, Seoul National University, Seoul, Korea.
\item Interdisciplinary Program in Bioinformatics, Seoul National University, Seoul, Korea.
\item Department of Electrical and Computer Engineering, University of California, San Diego, La Jolla, California, USA.
\\Correspondence should be addressed to S.Y. (sryoon@snu.ac.kr) or Y.K. (yhk@ucsd.edu).
\end{affiliations}

\begin{abstract}
Motivated by the need for fast and accurate classification of unlabeled nucleotide sequences on a large scale, we developed NASCUP, a new classification method that captures statistical structures of 
nucleotide sequences by compact context-tree models and universal probability from information theory.
NASCUP achieved BLAST-like classification accuracy consistently for several large-scale databases in orders-of-magnitude reduced runtime, and was applied to other bioinformatics tasks such as outlier detection and synthetic sequence generation.

\end{abstract}


Sequence classification plays a key role in various bioinformatics pipelines
by revealing the proximity and membership of a biological sequence to known sequence groups\cite{van2000virus, lu2005microrna, rfam, yap1993classification, tatusov2001cog}.
Expedited by new sequencing technologies, nucleotide sequence databases are
rapidly expanding at a rate that exceeds that of the technologies to handle the
bioinformatics around the sequences\cite{cochrane2012facing}.
Moreover, existing databases sometimes contain
mislabeled sequences that can potentially impair the identification accuracy significantly\cite{normand2018}. To address these challenges, we present NASCUP (http://github.com/nascup and \textbf{Supplementary Software}), an accurate and computationally efficient classification method that
is scalable for large and growing datasets and robust against mislabeling errors.

Common approaches to sequence classification can be broadly divided into two categories---alignment-based and model-based ones.
In alignment-based approaches, the class of the query sequence is determined by sequence-to-sequence comparison.
Alignment tools, including BLAST\cite{blast}, typically exhibit high accuracy, but are vulnerable to errors and often becomes time-consuming as the number of sequences increases.
Model-based approaches derive statistical models from each group of sequences and compare the query sequence to these models.
Sequence-to-model comparison is more scalable than sequence-to-sequence comparison, but it is often difficult to extract a model from a group of sequences that is statistically meaningful and does not overfit the data. By utilizing compact context-tree models along with the notion of universal probability from information theory, NASCUP delivers high accuracy comparable to the sequence-to-sequence comparison approaches, while providing robustness and scalability as a computationally efficient sequence-to-model comparison approach suitable for large-scale, expanding datasets.

Similar to most model-based sequence classification tools such as RDP\cite{methodRDP}, HMMER\cite{nhmmer}, and Phymm\cite{phymm}, the NASCUP pipeline consists of two stages (\textbf{\figurename~1a} and \textbf{\supfigurename~1}). In the first, model-building stage, NASCUP learns
the statistical structure of each nucleotide sequence
group in a database from the occurrence counts of all $k$-mers (substrings of length $k$) in the sequences
and builds a corresponding \emph{context-tree model}\cite{Willems--Shtarkov--Tjalkens1995}
(alternatively referred to as variable-order Markov models\cite{buhlmann1999} or probabilistic suffix trees\cite{Ron--Singer--Tishby1996}) that represents the data best
(\textbf{\figurename~1b} and \textbf{\supfigurename~2}).
Such context-tree models,
as reported for protein sequence classification\cite{Bejerano}
are simple enough for fast and scalable processing (as in $k$-mer count models of RDP), yet rich enough for accurate modeling of the data (as in hidden Markov models of HMMER or interpolated context models of Phymm).
In the second, classification stage of its pipeline,
NASCUP evaluates the likelihood of a test sequence under the context-tree model of each sequence group
and chooses the group that maximizes the likelihood.

In both model-building and classification stages, NASCUP utilizes the notion of \emph{universal probability}\cite{Algoet1992,Jiao--Permuter--Zhao--Kim--Weissman2013} from information theory. In a nutshell, universal probability approximates all probability distributions in a given class of models, and serves
as a close proxy to an unknown true probability distribution of given data without unnecessary overfitting. With theoretical performance guarantee and practical low-complexity
implementations, universal probability
has found many successful applications, including compression and prediction of sequential data of a priori unknown statistics. NASCUP measures how likely a sequence group fits a context-tree model
and how likely a test sequence fits the chosen context-tree model of a sequence group by evaluating the universal probabilities of the sequences.
The inference approach based on universal probability in particular and the information-theoretic principle in general has an additional benefit of having no tunable parameter\cite{keogh2004towards}
except the maximum depth of the context-tree models.

In order to demonstrate the classification accuracy and efficiency of NASCUP, we performed a set of comprehensive experiments on real sequence datasets from a variety of sources, organized on a functional or taxonomic basis with varying degrees of inter-group similarity (\textbf{\suptablename~1}).
NASCUP achieved
superb performance in accuracy and
speed among the five classification methods compared, consistently across the diverse set of seven datasets (\textbf{\figurename~1c}).
In terms of accuracy, NASCUP achieved
the highest average accuracy of 97.8\% (in terms of both arithmetic
and geometric means) among an expanded collection of thirteen alternative classification methods, which is trailed slightly by
the BLAST-based classification method (see \textbf{\suptablename~2}).
NASCUP also showed the highest accuracy both at the genus and species levels of the uncompacted SILVA dataset (\textbf{\suptablename~3}).
Except for NASCUP and BLAST, the classification accuracy varied significantly over the datasets.
In particular, HMMER was accurate on the functional RNA datasets but not on metagenomic microbial datasets, while RDP worked well on microbial datasets but showed unsatisfactory results on functional RNAs.
NASCUP, RDP, and USEARCH ran significantly (often by orders of magnitude) faster for most datasets (\textbf{\suptablename~4}).

To emulate the usage of classification tools
for a realistic environment in which sequence databases scale over time,
we prepared two types of expanding datasets---one by increasing the number of sequences for a fixed number of groups and the other by increasing the number of groups for a fixed number of sequences per group  (\textbf{\suptablename~5}).
For sequencewise expansion, the model building (first stage) time of NASCUP grew linearly as the number of sequences increased. The actual classification time, however, was affected only marginally since the second-stage classification operation is almost independent of the number of sequences in a group once the modeling has been completed (\textbf{\figurename~1d top}).
The total runtime of NASCUP (the sum of modeling and classification times) was lower than
the other four classification methods regardless of the data size (\textbf{\figurename~1e left} and \textbf{\suptablename~6}).

For groupwise expansion, the classification time of NASCUP grew as the number of groups (as well as the total number of sequences) increased. The modeling time did not increase since the model building procedure had to be performed only for newly added groups (\textbf{\figurename~1d bottom}). The performance of NASCUP was among the top under groupwise expansion, especially for very large databases (\textbf{\figurename~1e right} and \textbf{\suptablename~7}).
In both sequencewise and groupwise expansion experiments,
NASCUP was orders-of-magnitude faster than BLAST, the only method that achieved a comparable level of accuracy, across all database sizes.


We tested the robustness of NASCUP against
mislabeling errors in the sequence database,
which, in principle, exists in any real dataset labeled in the absence of the ground truth
and especially in pyrogenetic datasets.
We prepared a dataset with classification
errors at a rate ranging from 1\% to 20\%
and compared the classification accuracy
of five alternative methods
(\textbf{\figurename~1f} and \textbf{\supfigurename~5}).
The accuracy of NASCUP was robust with marginal
performance degradation even when 20\% of
the sequences in the database were mislabeled to arbitrary groups. RDP exhibited a similar level of robustness, whereas the performance of the other three methods degraded as the error rate increased.

As mentioned earlier, NASCUP computes the likelihood
of a test sequence belonging to each sequence group.
This numerical likelihood value provides additional
soft information that can augment the hard classification outcome. As the simplest application of
such likelihood values, we evaluated how the accuracy of NASCUP improves when it produces a ranked list of likely groups, instead of a single most likely group, for a given test sequence.
By increasing the list size, NASCUP achieved near-perfect accuracy
(\textbf{\figurename~2a}).
In particular, the ten most likely sequence groups produced by NASCUP included the correct group over 99\% of the time for all datasets.
Instead of producing a fixed number of candidate groups, NASCUP can produce a variable-size list according to the target accuracy, which can be estimated by the likelihood values and the Bayes rule (\textbf{\suptablename~8}).
The list of top candidate groups can be fed into subsequent bioinformatics pipelines (such as cross-validation by BLAST) or actual biological experiments on a far reduced set of groups than before preprocessed by NASCUP.

As another application of the likelihood value computed by NASCUP,
we performed one-class classification that identifies whether a test sequence belongs to a single target sequence group or not.
For NASCUP and HMMER, both of which build
generative models for the target group and provide the likelihood values of the test sequence.
We used normalized negative log likelihood value to distinguish member sequences in the target group from outliers. NASCUP showed a clear separation of outliers, which manifested in the much larger area under the precision--recall curve
(\textbf{\figurename~2b} and \textbf{\supfigurename~6}).

The combination of context-tree models
and universal probability in NASCUP finds
statistical generative models of nucleotide sequence groups
that are parsimonious and consequently are expected to better represent
the ground truth by Occam's razor
(\textbf{Supplementary Figs.~3} and~\textbf{4}).
In order to demonstrate the interpretive power of such generative models,
we generated synthetic sequences randomly according to four generative models---a combination of context-tree vs.\@ Markov models and universal vs.\@ maximum-likelihood probabilities---and measured how often these sequences looked real by classifying them using BLAST. As the maximum context size increased, BLAST
could classify the synthetic sequences generated
by NASCUP (context-tree models and universal probability) correctly with very high accuracy,
while the synthetic sequences generated by
the other three models failed to emulate
real sequences due to inadequate modeling
and overfitting
(\textbf{\figurename~2c} and \textbf{\supfigurename~7}).
Similar trends were observed when BLAST was replaced by other classification methods.

\section*{Acknowledgments}
This work was supported in part by the National Research Foundation (NRF) of Korea by Grants 2018R1A2B3001628 and 2014M3C9A3063541), and the Brain Korea 21 Plus Project 2018 Grant.




\newpage

\clearpage

\captionsetup{labelformat=empty}
\clearpage

\begin{figure}[!htbp]
\centering
\includegraphics[width=\textwidth]{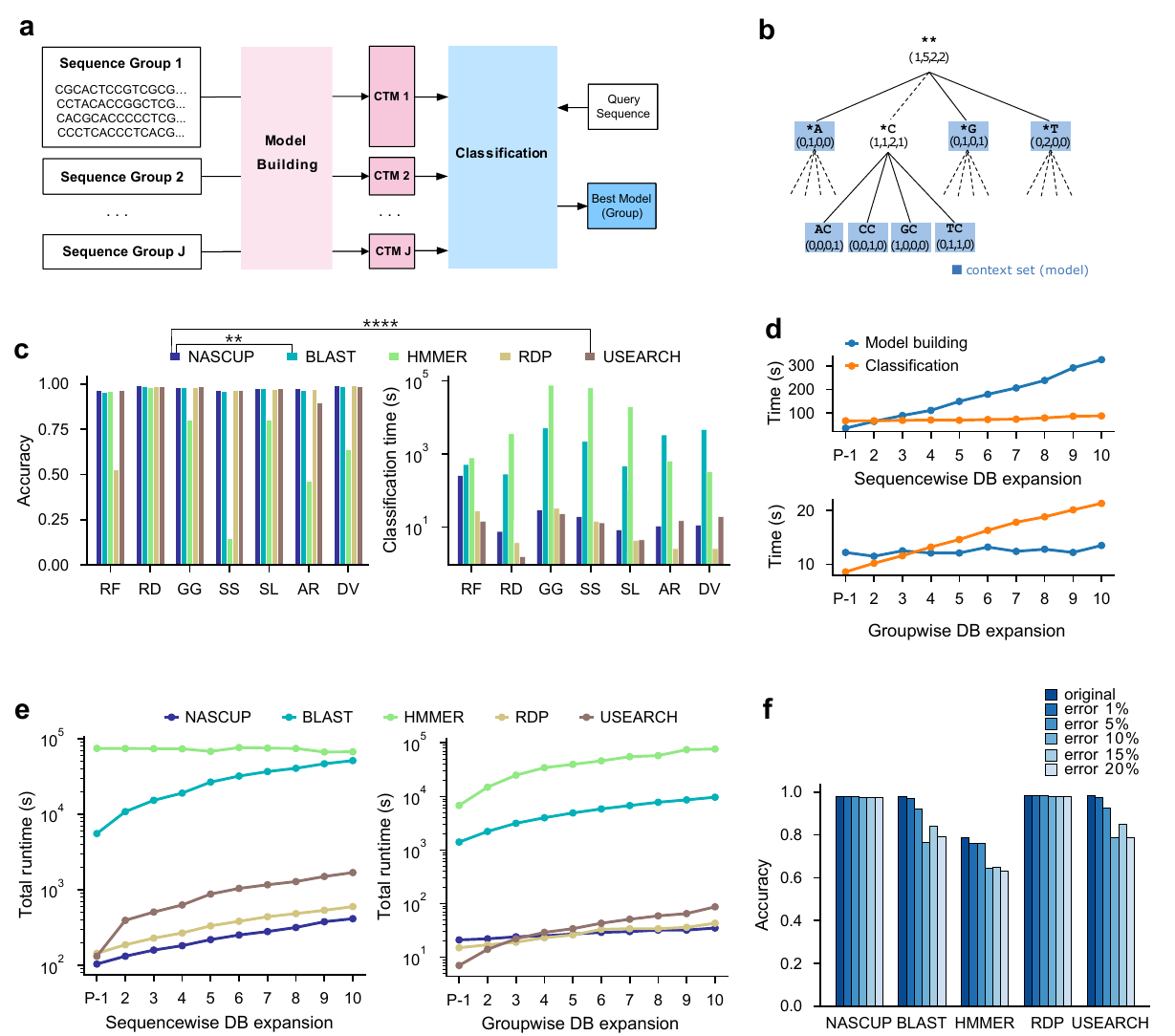}
\caption{} \label{fig:main}
\end{figure}
\vspace{-0.5in}
\noindent
\textbf{\figurename~\ref{fig:main}}
NASCUP pipeline and its classification performance.
(\textbf{a}) NASCUP pipeline consists of model building and classification stages.
(\textbf{b}) NASCUP builds context-tree models from sequence groups using universal probability.
(\textbf{c}) Accuracy and classification time (log scale) of \acl~and the four main alternatives (BLAST, HMMER, RDP, and USEARCH) on seven sequence datasets. Each value is the average of 10-fold cross validation. Statistical significance test (Wilcoxon signed-rank test) was performed between the best method and the next best method: NASCUP vs. BLAST, ** $P=2.0\times 10^{-3}$ and NASCUP vs. USEARCH, **** $P=2.0\times 10^{-16}$ (\textbf{\suptablename~9}).
(\textbf{d}) Model building and classification time of~\acl{} for sequencewise and groupwise expanding datasets. (\textbf{e}) Total runtime comparison of \acl~and the four main alternatives on expanding datasets.
 (\textbf{f}) Robustness of NASCUP and the four main alternatives for a mislabeled dataset (GG) as the error increases rate from 1\% to 20\%.

\clearpage
\begin{figure}[!htbp]
\centering
\includegraphics[width=\textwidth]{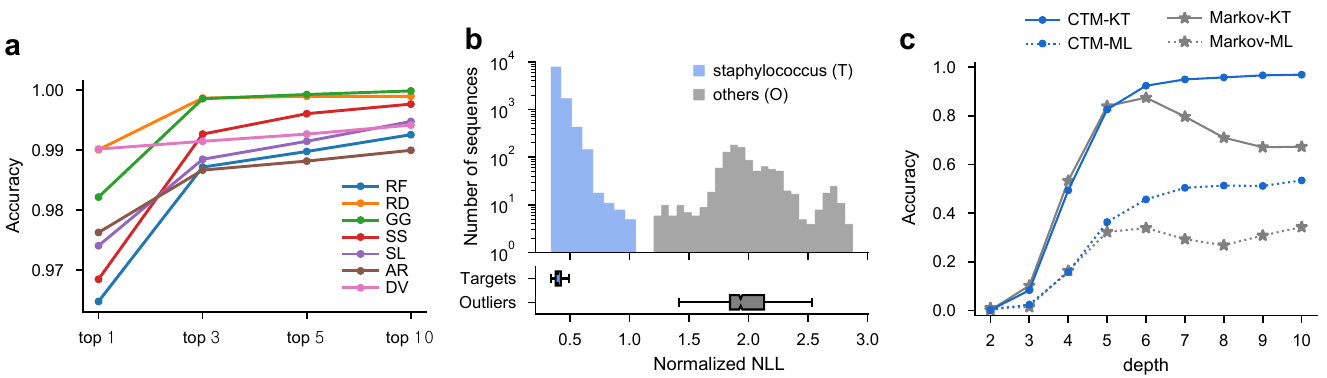}
\caption{}
\label{fig:sub}
\end{figure}
\vspace{-0.5in}
\noindent
\textbf{\figurename~\ref{fig:sub}}
Applications of the likelihood value and generative model of NASCUP.
(\textbf{a}) Classification accuracy of \acl~when lists of 1 to 10 candidates were formed.
(\textbf{b}) Outlier detection using NASCUP when 10,390
staphylococcus sequences were put together with 1,039 (10\%) outlier sequences chosen randomly from other genus-level groups. The distribution and box-plot of staphylococcus  and the outliers are plotted according to the normalized negative log-likelihood value.
(\textbf{c}) Synthetic sequence generation using NASCUP, whereby the quality of the generated sequence was measured by the classification accuracy of BLAST. In addition to context-tree models and Krichevsky--Trofimov universa probability estimates used in NASCUP,
3 other combinations using
Markov models and maximum likelihood (ML) probability estimates were also compared when
the maximum model depth varied from 2 to 10.

\clearpage 

\clearpage
\section*{ONLINE METHODS}

\newcommand{\ttX}{\texttt{X}}
\newcommand{\QKT}{Q^{\textrm{KT}}}
\newcommand{\QML}{Q^{\textrm{ML}}}
\newcommand{\sv}{{\bf s}}
\newcommand{\pv}{{\bf p}}
\newcommand{\xv}{{\bf x}}
\newcommand{\yv}{{\bf y}}
\newcommand{\Xv}{{\bf X}}
\newcommand{\Yv}{{\bf Y}}
\newcommand{\Sc}{{\cal S}}

\subsection{Data preparation.}
For basic classification experiments, we prepared seven datasets of nucleotide sequences (\textbf{\suptablename~1}): a function-based RNA family database (Rfam\cite{rfam}), taxonomy-based rRNA databases (RDP\cite{methodRDP,schloss2009introducing}, Greengenes\cite{desantis2006greengenes}, and SILVA-SSU/LSU\cite{quast2012silva}), and pyrosequencing databases (Artificial/Divergent\cite{quince}).
We excluded sequences containing symbols other than \texttt{ACGT}(\texttt{U}) and sequence groups of size less than ten for 10-fold cross-validation.
We compacted Greengenes and SILVA with cd-hit-est\cite{li2006cd} by 97\% similarity and limited the number of sequences in a group to 2,000.
The datasets thus obtained had diverse characteristics: the number of groups from 23 to 1,320, the sequence length from 20 to almost 5,000, and the average normalized intra-group pairwise sequence distance
from 0.08 to 0.33.
For scalability experiments, we generated two types of expanding datasets from the Greengenes database, each in ten phases (\textbf{\suptablename~5}). For the sequencewise expansion, we fixed a set of sequence groups and increased the number of sequences per group. For the groupwise expansion, we introduced a fixed number of new groups to an existing training set.

\subsection{Implementation and experimental setup.}
We implemented \acl~in \CC{}
(source code available at http://github.com/nascup)
and measured its performance in runtime (\ie, modeling time plus classification time)
and classification accuracy 
using a Linux machine (Ubuntu 12.04, 2.2 GHz Intel Xeon E5-4620, and 512 GB memory) without any parallelization.
We compared \acl{} with four main alternative methods (BLAST, HMMER, RDP, and USEARCH) as well as additional methods  (Phymm, gzip, UBLAST, caBLAST, BLAT, and three methods provided by QIIME\cite{qiime} --- Naive Bayes in QIIME-2, and UCLUST and Mothur in QIIME-1).
For BLAST and its variants based on sequence alignment (USEARCH, UBLAST, caBLAST, and BLAT), the class of a query sequence was determined by the best hit. For gzip, the class was determined by the smallest difference between the lengths of compressed representations of a sequence group and the group appended by the query sequence.
All command scripts of \acl~and other methods are provided as \textbf{\supnotename~2}.

\subsection{Markov and context-tree models.}
The simplest probabilistic model for nucleotide sequences is the independent and identical distribution (IID) model that assigns one of the four fixed probabilities $p_\ttA$, $p_\ttC$, $p_\ttG$, and $p_\ttT$ to each symbol, and computes the probability of the entire sequence as the product of those probabilities. More precisely, a sequence $\xv = \ttX_1\cdots\ttX_n$ with symbols $\ttA, \ttC, \ttG,$ and $\ttT$ respectively appearing $n_\ttA, n_\ttC, n_\ttG$, and $n_\ttT$ times has the probability
\begin{equation} \label{eq:prob_iid}
P(\xv) = p_\ttA^{n_\ttA}p_\ttC^{n_\ttC}p_\ttG^{n_\ttG}p_\ttT^{n_\ttT}.
\end{equation}
A $d$-th order Markov model introduces dependence across $d+1$ consecutive symbols (\ie, $k$-mers for $k = d+1$) by assigning
one of the probabilities $p_\ttA(\sv), p_\ttC(\sv), p_\ttG(\sv),$ and $p_\ttT(\sv)$
to the symbol $\ttX_i$ at position $i$
if the previous $d$ symbols $\ttX_{i-d}\cdots\ttX_{i-1}$, namely, the \emph{state}, is equal to $\sv$.
For example, a second-order Markov model assigns the probability $p_\ttG(\ttG\ttG)^2p_\ttA(\ttG\ttG)p_\ttC(\ttG\ttG)p_\ttG(\ttA\ttG)p_\ttG(\ttG\ttA)$
 to the sequence $\ttG\ttG|\ttG\ttA\ttG\ttG\ttG\ttC$ from the third position.
In general, a $d$-th order Markov model assigns the probability $P(\xv) = \prod_{i=d+1}^n p_{\ttX_i}(\ttX_{i-d}\cdots\ttX_{i-1})$ to the sequence $\xv = \ttX_1\cdots\ttX_d|\ttX_{d+1}\cdots\ttX_n$ from the $(d+1)$-st position. By parsing the sequence into subsequences by states, this probability can be equivalently expressed as
\begin{equation} \label{eq:prob_ctm}
P(\xv) = \prod_\sv P_\sv(\xv) = \prod_\sv p_\ttA(\sv)^{n_\ttA(\sv)}p_\ttC(\sv)^{n_\ttC(\sv)}p_\ttG(\sv)^{n_\ttG(\sv)}p_\ttT(\sv)^{n_\ttT(\sv)},
\end{equation}
where the products are over all states $\sv \in \{\nts\}^d$, $P_\sv(\xv)$ denotes the probability assigned to the subsequence of $\xv$ in state $\sv$ (that is, the symbols in $\xv$ that are preceded by $\sv$), and $n_\ttX(\sv)$ denotes the number of occurrences of symbol $\ttX \in \{\nts\}$ in that subsequence.
An IID model corresponds to a zeroth-order Markov ($1$-mer) model with the empty string as the only state, and that a $d$-th order Markov model can be decomposed as multiple IID models, each corresponding to one of the $4^d$ distinct states. A hidden Markov model used in HMMER\cite{nhmmer} is a generalization of
a Markov model by stochastically transforming a Markov model symbol-by-symbol.

A context-tree model (CTM)\cite{Willems--Shtarkov--Tjalkens1995} is both a specialization and generalization of a Markov model, in which states with a common suffix share the same probability assignments and thus are aggregated to form a \emph{context}. For example, the set $\{\ttA\ttA, \ttC\ttA, \ttG\ttA, \ttT\ttA\}$ consists of all possible states of a second-order Markov chain that has the common suffix $\ttA$.
This set is represented in shorthand notation $\ttS\ttA$, where $\ttS$ denotes ``any'' symbol.
Since the probability assignments are the same for all states in the context $\ttS\ttA$, the symbols preceded by $\ttA$ effectively follow a first-order Markov distribution. Since the effective Markov order varies from one context to another, a CTM is also referred to as a variable-order Markov model\cite{buhlmann1999}.
Each CTM is represented by a collection $\Sc$ of contexts that partition $\{\nts\}^d$
and parameters $\pv(\sv) = (p_\ttA(\sv), p_\ttC(\sv), p_\ttG(\sv), p_\ttT(\sv))$
associated with each context $\sv \in \Sc$. Using the contexts in $\Sc$ in as leaves
and merging contexts that share suffixes in a hierarchical manner, we can form a proper
(that is, each node has 0 or 4 children) suffix tree with root node $\ttS^d = \ttS\cdots\ttS$ ($d$ times).
For example, the contexts $\ttS\ttA, \ttA\ttC, \ttC\ttC,\ttG\ttC, \ttT\ttC, \ttS\ttG, \ttS\ttT$
partition $\{\nts\}^2$ and form a suffix tree (\textbf{\supfigurename~2}).
Consequently, a CTM is equivalently represented by a suffix tree along with probability assignments
on its leaves, and thus is referred to as a probabilistic suffix tree model\cite{Ron--Singer--Tishby1996} as well.
A $d$-th order Markov model can be viewed as a CTM on a perfect suffix tree that has $\{\nts\}^d$ as
its $4^d$ leaves. A typical CTM of depth $d$ consists of a fewer number of contexts (and corresponding probability parameters) than a $d$-th order Markov model, providing a more succinct representation of the data.
A CTM can be further generalized by allowing states to be merged to a context at any position,
\eg, $\ttA\ttS = \{\ttA\ttA, \ttA\ttC, \ttA\ttG, \ttA\ttT\}$.
An interpolated context model used in Phymm\cite{phymm} is a mixture of such generalized CTMs.
NASCUP uses only suffix trees since this restriction does not incur any empirical performance loss compared to CTMs, and it allows computationally efficient comparison among all suffix trees.

\subsection{Maximum likelihood and universal probability estimates.}
When the parameters $p_\ttA, p_\ttC, p_\ttG, p_\ttT$ of an IID model are not known, the most naive approach to estimating the true probability $P(\xv)$ of a given sequence $\xv$ is to find the parameters that maximize the sequence probability expression in~\eqref{eq:prob_iid} and then to compute the probability of the entire sequence using these parameter estimates. It can be easily verified that
the empirical frequencies $(p_\ttA, p_\ttC, p_\ttG, p_\ttT) = (n_\ttA/n, n_\ttC/n, n_\ttG/n, n_\ttT/n)$ maximize \eqref{eq:prob_iid} for any sequence $\xv$
with symbol counts
$n_\ttA,n_\ttC,n_\ttG,n_\ttT$ and length
$n = n_\ttA + n_\ttC + n_\ttG + n_\ttT$.
The resulting \emph{maximum likelihood (ML) estimate} of the true probability is
\begin{equation} \label{eq:prob_ml}
\QML(\xv) = \biggl(\frac{n_\ttA}{n}\biggr)^{n_\ttA}\biggl(\frac{n_\ttC}{n}\biggr)^{n_\ttC}\biggl(\frac{n_\ttG}{n}\biggr)^{n_\ttG}\biggl(\frac{n_\ttT}{n}\biggr)^{n_\ttT}.
\end{equation}
Note that $\QML(\xv)$ is a function only of the symbol count vector $\bn = (n_\ttA,n_\ttC,n_\ttG,n_\ttT)$.
Hence, with a slight abuse of notation, we will write the righthand side (RHS) of~\eqref{eq:prob_ml}
as $\QML(\bn)$.
More generally, for a depth-$d$ CTM on a given context set $\Sc$ with unknown parameters $\pv(\sv)$, $\sv \in \Sc$,
the ML estimate $\QML(\xv)$ of the true probability $P(\xv) = \prod_\sv P_\sv(\xv)$ (from the $(d+1)$-st position) as in~\eqref{eq:prob_ctm}
is the product of the ML estimates of subsequence probabilities $P_\sv(\xv)$, namely,
\begin{equation} \label{eq:prob_ml_ctm}
\QML(\xv) = \prod_{\sv \in \Sc} \QML(\bn(\sv)),
\end{equation}
where $\QML(\bn(\sv))$ denotes the ML estimate of an IID probability in~\eqref{eq:prob_ml} evaluated with
the count vector $\bn(\sv) =  (n_\ttA(\sv),n_\ttC(\sv),n_\ttG(\sv),n_\ttT(\sv))$ for context $\sv$.
The ML estimate $\QML(\xv)$ overfits the given data $\xv$ by discounting the symbols that did not appear in it. In fact, $\QML(\cdot)$ is not a valid probability assignment since the sum of the estimated probabilities $\QML(\xv)$ over all sequences $\xv \in \{\nts\}^n$ is greater than 1.

As an alternative, NASCUP relies on the notion of \emph{universal probability}\cite{Algoet1992,Jiao--Permuter--Zhao--Kim--Weissman2013} in information theory
that is chosen independent of the data $\xv$ and is close to all unknown probability models in a given class. The most basic example of universal probability is
the \emph{Krichevski--Trofimov (KT) estimate}\cite{Krichevsky--Trofimov1981}
for IID models that assigns the sequence probability
\begin{align}
\QKT(\xv)
&= \int p_\ttA^{n_\ttA}p_\ttC^{n_\ttC}p_\ttG^{n_\ttG}p_\ttT^{n_\ttT} f(p_\ttA, p_\ttC, p_\ttG, p_\ttT)
dp_\ttA dp_\ttC dp_\ttG dp_\ttT, 
\label{eq:prob_kt}
\end{align}
where $f(p_\ttA, p_\ttC, p_\ttG, p_\ttT)$ is the Dirichlet prior on the quaternary probability simplex with parameters $1/2,1/2,1/2,1/2$.
As a Dirichlet mixture of IID models, the KT estimate is a valid probability assignment (that is, $\QKT(\xv) \ge 0$ for every $\xv$ and $\sum_\xv \QKT(\xv) = 1$).
Moreover, $\QKT$ is \emph{uniformly} close to every IID probability model $P$ on quaternary
sequences of length $n$
in the sense that both the relative entropy (Kullback--Leibler divergence)\cite{Cover--Thomas2006}
\[
D(P\|\QKT) = \sum_{\xv \in \{\nts\}^n}
P(\xv) \log \frac{P(\xv)}{\QKT(\xv)}
\]
and the maximum log likelihood ratio
\[
\max_{\xv} \log \frac{P(\xv)}{\QKT(\xv)}
\]
are upper bounded by $(3/2) \log n$ plus uniform constants independent of $P$,
which vanishes when normalized by the sequence length $n$ and is essentially tight
as no other probability estimate can approximate all IID probability models uniformly closer\cite{Rissanen1984,Shtarkov1987}.

Since the Dirichlet distribution is the conjugate prior for the parameters of an IID model, the KT probability estimate has the predictive ``add-half'' formula for the conditional probability
\begin{equation} \label{eq:add_half}
\QKT(\ttX_{i+1} = \ttX | \ttX_1,\ldots,\ttX_i)
= \frac{i_\ttX + 1/2}{i + 2}, \quad \ttX \in \{\ttA,\ttC, \ttG, \ttT\},\, i = 0,1,2,\ldots,
\end{equation}
when the sequence $\ttX_1\cdots\ttX_i$ has symbol counts $i_\ttA, i_\ttC, i_\ttG, i_\ttT$.
Applying
this predictive estimate sequentially, the KT probability estimate of the entire length-$n$ sequence $\xv$ in~\eqref{eq:prob_kt} can be expressed as
\begin{equation}
\QKT(\xv)
= \frac{\prod_{\ttX \in \{\nts\}} \prod_{i_\ttX=1}^{n_\ttX} (i_\ttX - 1/2) }
{ \prod_{i=1}^n (i + 1)}
= \frac{\prod_{\ttX \in \{\nts\}}\Gamma(n_\ttX + 1/2)}
{\pi^2 \Gamma(n + 2) }, \label{eq:prob_kt2}
\end{equation}
where $\Gamma(\cdot)$ is the standard Gamma function.
As with the ML estimate, the KT estimate is a function of $\xv$ only through
the symbol count vector $\bn = (n_\ttA,n_\ttC,n_\ttG,n_\ttT)$ and hence we will
write the RHS of~\eqref{eq:prob_kt2} as $\QKT(\bn)$.
Further extending the predictive estimate in~\eqref{eq:add_half},
we can express the conditional probability of a sequence $\xv$ with symbol count vector $\bm$
given a preceding sequence $\yv$ with symbol count vector $\bn$
as
\begin{equation} \label{eq:cond_kt}
\QKT(\xv|\yv)
= \frac{\QKT(\yv,\xv)}{\QKT(\yv)} = \frac{\QKT(\bm + \bn)}{\QKT(\bn)}.
\end{equation}

Paralleling~\eqref{eq:prob_ctm} and~\eqref{eq:prob_ml_ctm}, we can generalize the KT estimate
to a CTM on a given context set $\Sc$ with unknown parameters $\pv(\sv)$. In this case, the KT estimate
of the probability of the sequence $\xv$ with symbol count vectors $\bn(\sv)$, $\sv \in \Sc$, is
\begin{equation}  \label{eq:prob_kt_ctm}
\QKT(\xv) = \prod_{\sv \in \Sc} \QKT(\bn(\sv))
\end{equation}
and the KT estimate of the conditional probability of $\xv$ with symbol count vectors $\bm(\sv)$ given $\yv$ with symbol count vectors $\bn(\sv)$ is
\begin{equation} \label{eq:prob_kt_ctm_cond}
\QKT(\xv|\yv) = \prod_{\sv \in \Sc} \frac{\QKT(\bm(\sv)+\bn(\sv))}{\QKT(\bn(\sv))}.
\end{equation}
As in the IID case, the KT probability estimate is universal in the sense that $\QKT$ is uniformly close to all CTMs on the given context set $\Sc$.

\subsection{Model building.}
Given a collection of sequence groups in a database, NASCUP models each sequence group by a CTM
of an unknown context set $\Sc$ and unknown parameters $\pv(\sv)$, $\sv \in \Sc$. NASCUP estimates
the probability of such an unknown CTM by selecting the ``maximum likelihood'' context set $\Sc^*$ based on the sequences $\yv$ in the group. Since the parameters are unknown, NASCUP evaluates the universal probability
$\QKT(\yv; \Sc)$ for all possible $\Sc$, and chooses $\Sc^*$ that attains the maximum.
This context tree selection procedure of NASCUP is intimately related
to the model selection method of the \emph{minimum description length principle}\cite{Rissanen1978,Quinlan--Rivest1989}
and other information criteria used in statistics and information theory.
The resulting context set $\Sc^*$ and the count vectors are used for the subsequent classification stage.
\textbf{\supfigurename~1} represents the overall flow of the \acl.

The detail of the model-building stage is as follows.
NASCUP initially counts the nucleotide symbols \nts{}
for all sequences in the group
that follow each of length-$d$ contexts $\sv \in \{\nts,\ttS\}^d$
and forms count vectors $\bn(\sv)$.
These count vectors can be calculated by merging bottom--up from the leaves of the perfect suffix tree (the $d$-th order Markov model) to the root (the IID model) iteratively. As an example, for $d = 2$, $\bn(\ttS\ttA) = \bn(\ttA\ttA) + \bn(\ttC\ttA) + \bn(\ttG\ttA) + \bn(\ttT\ttA)$
and $\bn(\ttS\ttS) = \bn(\ttS\ttA) + \bn(\ttS\ttC) + \bn(\ttS\ttG) + \bn(\ttS\ttT)$.
NASCUP then finds the context set $\Sc$ that maximizes
the KT estimate in~\eqref{eq:prob_kt_ctm}, which can be viewed as a proxy for all unknown CTMs on $\Sc$.
This maximization can be performed efficiently in a recursive manner\cite{Willems--Shtarkov--Tjalkens1993, Nohre1994} by starting with the root in the perfect suffix tree
in a depth-first manner and computing for each node $\sv \in \{\nts,\ttS\}^d$ in the tree
\begin{equation}
Q(\sv) = \max \biggl\{ \QKT(\bn(\sv)),\prod_{\sv': \textrm{ children of $\sv$}} Q(\sv') \biggr\}.
\label{eq:maximizing}
\end{equation}
If $\sv$ is a leaf (at depth $d$), then $Q(\sv) = \QKT(\bn(\sv))$. As an example,
for $d = 2$,
\begin{align*}
Q(\ttS\ttS)
&= \max\{ \QKT(\bn(\ttS\ttS)),\,
Q(\ttS\ttA)Q(\ttS\ttC)Q(\ttS\ttG)Q(\ttS\ttT)\} \\
\intertext{and}
Q(\ttS\ttA) &= \max\{ \QKT(\bn(\ttS\ttA)), \,
\QKT(\bn(\ttA\ttA))\QKT(\bn(\ttC\ttA))\QKT(\bn(\ttG\ttA))\QKT(\bn(\ttT\ttA)) \}.
\end{align*}%
A branch that does not attain the maximum is pruned. A tie between a parent node and its children is broken against branching for the sparsity of the resulting model.
Upon completion of this maximizing and pruning step,
At the end of model building for each sequence group, NASCUP produces a valid context set $\Sc^*$
and symbol count vectors $\bn(\sv)$ for all contexts $\sv \in \Sc^*$.

\subsection{Classification.}
The notion of universal probability plays a pivotal role in the classification stage as well.
If the true CTM $P_j$ for sequence group~$j$ in the database were known,
then the maximum likelihood
classifier would compute $P_j(\xv)$ of the query sequence $\xv$
for all groups~$j$ and select the group~$j^*$
that maximizes the likelihood. As in the model-building stage, NASCUP relies on the first principle of using the universal probability when the true probability is unknown, and computes the maximum
using the universal probability instead of $P_j$.
More concretely, NASCUP first
generates count vectors $\bm(\sv)$ from the query sequence $\xv$ for all contexts $\sv \in \{\nts,\ttS\}^d$ bottom--up.
For each sequence group~$j$ with context tree $\Sc_j^*$ and count vectors $\bn(\sv)$, $\sv \in \Sc_j^*$,
NASCUP then uses~\eqref{eq:prob_kt_ctm_cond} to compute
the KT estimate  $\QKT(\xv | \yv_j)$ of the conditional probability
of the query sequence $\xv$ given the existing sequences $\yv_j$ in group~$j$.
Assuming that the correct context tree $\Sc_j^*$ was found,
this KT estimate $\QKT(\xv | \yv_j)$ is universal and thus
is uniformly close to the true conditional probability $P_j(\xv | \yv)$, which is in turn uniformly close
to $P_j(\xv)$ due to the Markov property of the CTM.
Hence, we can perform ML classification approximately
without knowing the true probability distributions.
NASCUP thus compares $\QKT(\xv | \yv_j)$ over all sequence groups~$j$
and selects the one with the maximum KT conditional probability.
The idea of using universal probability in sequence classification
traces back to the information theory literature\cite{Ziv1988,Gutman1989} and NASCUP extends it to
CTMs and associated universal probability estimates.
Note that the measure, $\log 1/P_j(\bx|\by_j)$, can be interpreted as
the code length for lossless compression of the sequence\cite{Cover--Thomas2006}
under the probability model $P_j$.
In this sense, NASCUP can be viewed as a refinement of
the nucleotide sequence classification methods based on compression\cite{loewenstern1995dna,kocsor2006application} in that the code length from NASCUP is essentially optimal and NASCUP provides the probability model itself instead of the code length.

\subsection{Scores, ranking, and multi-candidate classification.}
For each sequence group $j = 1,\ldots, J$ in the database, NASCUP computes the estimate
$\QKT(\xv | \yv_j)$ of the likelihood that the query sequence is generated from group~$j$.
This likelihood estimate serves as a score for each group,
which can be rank-ordered to form a small list of candidate groups of the query sequence
(\textbf{\suptablename~8}).
Such a candidate list can boost the accuracy of classification, for example, as an input to a slower yet more accurate classifier, or even as a focused target for biological experiments.
The simplest approach to forming a list is to sorting all $J$ groups in the database by the likelihood and choosing the top $K$ groups for a fixed number $K$. Alternatively, the list size $K$ can be adjusted adaptively by estimating the overall accuracy of the list. By the Bayes rule, under the uniform prior on the sequence groups,
the posterior probability that the sequence $\xv$ belongs group~$j$ is approximately
\[
\frac{\QKT(\xv | \yv_j)}{\sum_{j = 1}^J \QKT(\xv | \yv_j)}.
\]
The sum of these posterior probabilities of the candidates in a list provides an estimate of the classification accuracy, which can be used to control the size of the list.
Note that this approach can incorporate an arbitrary prior on the sequence groups.

\subsection{Outlier detection.}
The CTM that NASCUP finds in the model-building stage
and
the resulting KT estimate of the conditional probability in~\eqref{eq:prob_kt_ctm_cond}
can be utilized beyond classification of query sequences.
Suppose that there are a few outliers in a sequence group. Since the symbol counts used by NASCUP reflects the statistical behavior of the entire group, the generated model is rather immune to a small number of outliers and other errors in the data.
Once the model is built, we can detect outliers within the group by evaluating the universal probability $\QKT(\xv|\yv)$ trained by all sequences $\yv$ in the group with each individual sequence $\xv$ in the group. We measure the degree of conformance to the model of each sequence $\xv$ of length $m$ by
its normalized negative log-likelihood (NLL)
\[
\frac{1}{m} \log \frac{1}{\QKT(\xv|\yv)}.
\]
The smaller the NLL value is, the better the conformance to the model is.
Conversely, the larger the value, the greater the difference from the model, indicating
a high likelihood of being an outlier (\textbf{\supfigurename~6}).

\subsection{Synthetic sequence generation.}
Let $d$ be the depth of the context tree and $l$ be the average length of the sequences $\yv$ in a sequence group. First, we generate the starting string $\ttX_1\cdots\ttX_d$ of length~$d$ by copying the most frequent starting string of the same length among the existing sequences in the group.
Using a suffix of $\ttX_1\cdots\ttX_d$ as the context $\sv$, we generate the next symbol $\ttX_{d+1}$ according to
\[
Q^{KT}(\ttX_{d+1} = \ttX | \ttX_1, \ldots,\ttX_d, \yv) = \frac{n_\ttX(\sv) + 1/2}{n(\sv) + 2}
\]
as in~\eqref{eq:add_half}. Subsequently, we generate $\ttX_i$, $i = d+2,\ldots, l$, each according to
a similar predictive probability estimate with a suffix of $\ttX_{i-d}\cdots\ttX_{i-1}$ as the context
and the corresponding count vector from the existing sequences $\yv$ as well as the preceding symbols
$\ttX_{d+1}\cdots\ttX_{i-1}$. This sliding-window sequence generation procedure is an extension of Polya's urn process in the standard Bayesian statistics to a CTM. Due to the conjugacy of the Dirichlet prior, the distribution of the generated sequence $\xv$ (from the $(d+1)$-st position and on) is equivalent to a CTM with \emph{random} parameters $\pv(\sv)$ drawn from the Dirichlet prior with parameters $n_\ttA(\sv) + 1/2, n_\ttC(\sv)+1/2, n_\ttG(\sv)+1/2, n_\ttT(\sv)+1/2$ for each $\sv \in \Sc$.
To ascertain the quality of synthetic sequences in our procedure,
we generated synthetic sequences from RF, RD, and GG datasets, one per group, and classified them
using BLAST (\textbf{\supfigurename~7}).



\subsection{Statistical significance.}
We performed Wilcoxon signed-rank tests in the SciPy library on normalized and unnormalized aggregate datasets as well as individual datasets (\textbf{\suptablename~9}).
The normalized aggregate dataset was formed by combining the seven datasets in
\textbf{\suptablename~1}. In order to balance the impacts of the individual datasets on the aggregate, we
drew bootstrap samples of the same size from each. The sample size of 2,900 was used, which is
one tenth of the average size of the component datasets excluding the smallest (RD) and the largest (RF).
We used the averaged p-value over 100 signed-rank tests repeated on
independently generated bootstrap samples.
For the unnormalized aggregate dataset, we used all samples in the seven datasets without normalization  of their sizes.
All pairs of the five main classification methods (NASCUP, BLAST, HMMER, RDP, USEARCH)
were compared for statistical significance.

\subsection{Other experiments.}
Additional experimental results, including design alternatives to CTM and universal probability used in NASCUP, are described in \textbf{\supnotename~1}.
%
%



\bibliographystyle{naturemag}
\bibliography{reference}

\clearpage
\newpage
\includepdf[pages=1-last]{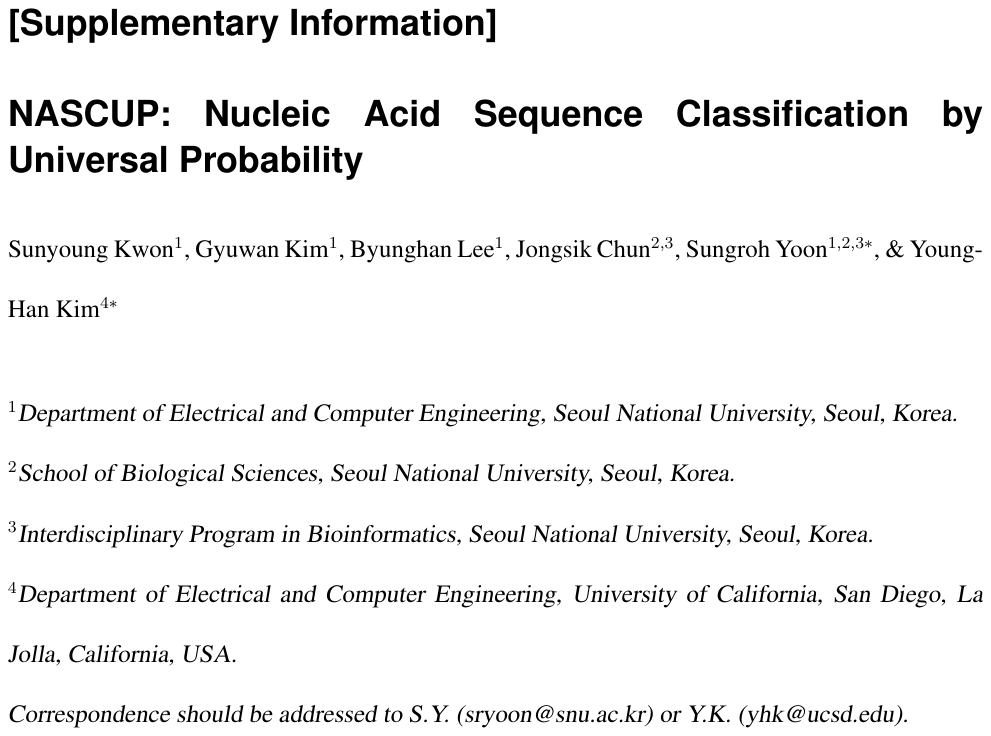}


\end{document}


\maketitle

\begin{affiliations}
\item Department of Electrical and Computer Engineering, Seoul National University, Seoul, Korea.
\item School of Biological Sciences, Seoul National University, Seoul, Korea.
\item Interdisciplinary Program in Bioinformatics, Seoul National University, Seoul, Korea.
\item Department of Electrical and Computer Engineering, University of California, San Diego, La Jolla, California, USA.
\\Correspondence should be addressed to S.Y. (sryoon@snu.ac.kr) or Y.K. (yhk@ucsd.edu).
\end{affiliations}

\clearpage

\captionsetup{labelformat=empty}
\renewcommand{\figurename}{Supplementary Fig.}
\renewcommand{\tablename} {Supplementary Table.}

\newcommand{\mylinespacing}{1.3}
\newcommand{\mytablespacing}{1.1}

\newcommand{\supspace}{\vspace{15pt}}

\section*{Supplementary Information}

\begin{spacing}{1.5}
\begin{tabular}{ll}
\supnotename~\ref{note:result} & Other experiments \\
\supnotename~\ref{note:script} & Command scripts and version information used in the experiments \\
\\
\supfigurename~\ref{fig:overall_flow} & Overall flow of \acl \\
\supfigurename~\ref{fig:context_graph} & Context-tree model \\
\supfigurename~\ref{fig:varaccu} & Accuracy of \acl~ variants \\
\supfigurename~\ref{fig:varleaf} & The ratio of the number of leaf nodes \\
\supfigurename~\ref{fig:noise} & Performance degradation from mislabeling errors \\
\supfigurename~\ref{fig:outlier} & Outlier detection with normalized NLL\\
\supfigurename~\ref{fig:gen} & Synthetic sequence generation and its quality assessment\\
\\
\suptablename~\ref{tab:data} & Details of the dataset used in the experiments\\
\suptablename~\ref{tab:result} & Classification accuracy comparison\\
\suptablename~\ref{tab:silva} & Experimental results on large-scale SILVA (SSS, LSU) datasets\\
\suptablename~\ref{tab:time} & Total runtime (classification time) comparison \\
\suptablename~\ref{tab:data_sc} & Details of the expanding datasets used for scalability experiments\\
\suptablename~\ref{tab:bggs_time} & Runtime in the sequencewise expansion datasets \\
\suptablename~\ref{tab:bggf_time} & Runtime in the groupwise expansion datasets \\
\suptablename~\ref{tab:recall} & Accuracy comparison from candidate extraction \\
\suptablename~\ref{tab:stat} & Statistical significance tests\\
\end{tabular}
\end{spacing}

\begin{spacing}{\mylinespacing}

\clearpage
\section*{\supnotename~\ref{note:result}}
\supspace
\begin{spacing}{\mylinespacing}
\begin{listing}[htbp]
\caption{ }
\label{note:result}
\end{listing}

\vspace{-0.5in}
\noindent\textbf{Comparison of \acl~and its design alternatives.}
Universal probability and context-tree models are two key features of~\acl. In order to demonstrate their combined benefit, we compared classification accuracy among a few design alternatives by varying the model depth (\textbf{\supfigurename~\ref{fig:varaccu}}).
As an alternative to a CTM, a Markov model provides a baseline 
before pruning context trees based on recursive likelihood maximization.
For probability assignment, the maximum likelihood (ML) estimator
provides a naive alternative to the KT estimator
in assigning the conditional probability in each context.
%
%
%
%
Since the number of possible contexts grows exponentially as the model depth increases, 
the occurrence of each context becomes sparse with a restricted number of training data.
Consequently, the Markov model and ML estimator are prone to overfitting in general.
In our accuracy comparison experiments, both Markov and context-tree models with ML estimator
performed rather poorly.
While the performance advantage of CTMs over Markov models was not very pronounced,
the numbers of leaves in CTMs did not increase exponentially as those in Markov models and their ratios dropped rapidly as the depth increased
(\textbf{\supfigurename~\ref{fig:varleaf}}). 
Moreover, CTMs using KT estimator always have fewer leaf nodes than CTMs using ML estimator.
In conclusion, the combination of KT and CTM used in NASCUP was the most parsimonious.
This sparsity can be interpreted as being closer to the ground truth in principle (which was cross-examined by the synthetic sequence generation experiment), but also leads to
faster and more efficient classification as a practical benefit.

\noindent\textbf{Comparison of NASCUP with twelve alternative classification methods.}
Including the four main alternatives, we compared \acl~with twelve classification methods in terms of accuracy and running time (\textbf{\suptablename{}s~\ref{tab:result} and~\ref{tab:time}}). Regardless of the diversity of dataset, \acl~showed a high level of accuracy consistently and achieved the best performance in term of the average and geometric mean accuracy. 
In particular, \acl~showed the highest accuracy on RF dataset, which was the most difficult dataset to classify due to its largest number of classes and widest intra-group sequence distance (AIGD). More than half of the thirteen classification methods (UBLAST, BLAT, caBLAST, RDP, gzip, UCLUST, and Mothur) exhibited unsatisfactory results of below 80\% of accuracy on RF. 
\acl~and BLAST maintained the accuracy of above 95\% across all datasets considered, whereas the performance of the other methods varied often significantly from dataset to dataset.
\noindent \acl, RDP, USEARCH, and QIIME package based methods ran significantly (often by orders of magnitude) faster for most of the datasets. 

\noindent\textbf{Runtime in the sequencewise and groupwise expanding datasets.}
With advances in sequencing technologies, sequence databases are constantly being updated and expanded. 
To ascertain the desired scalability of NASCUP in such environments,
we performed classification experiments on two types of database expansion---sequencewise and groupwise expanding datasets (\textbf{\suptablename~\ref{tab:bggs_time} and~\ref{tab:bggf_time}}).
In the sequencewise expansion experiment, the classification time of alignment-based methods increased proportionally to the number of sequences in the database, while the classification time of the model-based methods including NASCUP were affected by the number of groups. As the number of sequences in a group increased, the time difference became wider. 
In the groupwise expansion experiment, model-based methods require separate model building time only for newly added groups. The larger the database was, the greater gain in  the classification time was observed for model-based methods.
NASCUP was a top scalable method in the sequencewise and groupwise expansion experiments while maintaining the accuracy comparable to BLAST.

\end{spacing}

\clearpage
\section*{\supnotename~\ref{note:script}}
\supspace
\begin{spacing}{\mylinespacing}

\noindent\textbf{Command scripts and version information used in the experiments.}
We used FASTA formatted nucleotide sequence datasets, train.fasta for training and test.fasta for testing. The train.fasta contains sequences of the entire groups in a single file, and each of train\_i.fasta contains sequences corresponding to a specific $i$-th individual group as requested by some programs.
The test.fasta contains all test sequences in a single file, but each of test\_i.fasta contains only a specific $i$-th single sequence as requested by some programs.
\\

\end{spacing}

\begin{listing}[htbp]
\caption{ }
\label{note:script}
\end{listing}

\begin{spacing}{1.1}
\vspace{-0.3in}

\noindent NASCUP
\vspace{-0.3in}
\begin{minted}[fontsize=\footnotesize]{text}
# version 201803
nascup_build -i train.fasta -m kt -d 6 -o train.model
nascup_scan -c train.model -i test.fasta -m kt -o output
\end{minted}

\noindent BLAST
\vspace{-0.3in}
\begin{minted}[fontsize=\footnotesize]{text}
# version 2.2.25
formatdb -p F -i train.fasta
blastall -a 1 -p blastn -d train.fasta -i test.fasta -o output -m 8
\end{minted}

\noindent USEARCH
\vspace{-0.3in}
\begin{minted}[fontsize=\footnotesize]{text}
# version 8.1.1861
usearch -usearch_local test.fasta -db train.fasta --blast6out output \
-strand plus -id 0.8 -threads 1 -top_hit_only
\end{minted}

\noindent UBLAST
\vspace{-0.3in}
\begin{minted}[fontsize=\footnotesize]{text}
# version 8.1.1861
usearch -makeudb_ublast train.fasta -output train.udb
usearch -ublast test.fasta -db train.udb -blast6out -evalue 1e-9 \
-strand plus -threads 1 -top_hit_only -accel 0.8
\end{minted}

\noindent BLAT
\vspace{-0.3in}
\begin{minted}[fontsize=\footnotesize]{text}
# version 35
# generate train_rename.fasta from train.fasta
blat -noHead train_rename.fasta test.fasta output -out=blast8
\end{minted}

\noindent caBLAST
\vspace{-0.3in}
\begin{minted}[fontsize=\footnotesize]{text}
# version 1.2.1
cablast-compress TRAIN train.fasta
for j in idx_of_test_sequence :
	cablast-search TRAIN test_j.fasta
\end{minted}

\noindent HMMER
\vspace{-0.3in}
\begin{minted}[fontsize=\footnotesize]{text}
# version 3.1b2
rm total_hmm
for i in idx_of_group :
    clustalo --full -i train_i.fasta -o train_i.msa --outfmt=st --force
    hmmbuild --cpu 1 -o tmp --dna train_i.hmm train_i.msa
    cat train_i.hmm >> total_hmm
hmmpress total_hmm
nhmmscan --cpu 1 -o tmp --tblout output total_hmm test.fasta
\end{minted}

\noindent RDP
\vspace{-0.3in}
\begin{minted}[fontsize=\footnotesize]{text}
# version 2.11
# generate train_rdp.fasta and train_taxid.txt from train.fasta
python train_rename.py train.fasta train_rdp.fasta train_taxid.txt"
java -Xmx1g -jar rRNAClassifier.properties train -o part/RDP \
-s train_rdp.fasta -t train_taxid.txt
java -Xmx1g -jar classifier.jar classify -t rRNAClassifier.properties \
-o output test.fasta
\end{minted}

\noindent Phymm
\vspace{-0.3in}
\begin{minted}[fontsize=\footnotesize]{text}
# version 4.0
for i in idx_of_group:
    build-icm -d 6 -w 12 -p 1 test_i.icm < train_i.fasta
    simple-score -N test_i.icm < test.fasta > output
\end{minted}

\noindent gzip
\vspace{-0.3in}
\begin{minted}[fontsize=\footnotesize]{text}
# version (GNU tar) 1.26
for i in idx_of_group:
    tar -czf train_i.gzip train_i.fasta
    for j in idx_of_test_sequence :
        cat train_i.fasta test_j.seq > train_ij.fasta
        tar -czf train_ij.gzip train_ij.fasta
\end{minted}

\newpage
\noindent QIIME2
\vspace{-0.3in}
\begin{minted}[fontsize=\footnotesize]{text}
# version QIIME 2
# generate train_rename.fasta from train.fasta
# generate train_taxid.txt from train.fasta
qiime tools import --type 'FeatureData[Sequence]' --input-path train_rename.fasta \
--output-path  train_rename.qza
qiime tools import --type 'FeatureData[Taxonomy]' \
--source-format HeaderlessTSVTaxonomyFormat \
--input-path train_taxid.txt --output-path train_taxid.qza
qiime tools import --type 'FeatureData[Sequence]' --input-path test.fasta \
--output-path test.qza
qiime feature-classifier fit-classifier-naive-bayes \
--i-reference-reads train_rename.qza --i-reference-taxonomy train_taxid.qza \
--o-classifier classifier.qza
qiime feature-classifier classify-sklearn --i-classifier classifier.qza \
--i-reads test.qza -o-classification qiime2.qza
qiime tools export qiime2.qza --output-dir result
\end{minted}

\noindent UCLUST
\vspace{-0.3in}
\begin{minted}[fontsize=\footnotesize]{text}
# version QIIME 1
# generate train_rename.fasta from train.fasta
# generate train_taxid.txt from train.fasta
assign_taxonomy.py -i test.fasta -r train_rename.fasta -t train_taxid.txt -m uclust
\end{minted}

\noindent Mothur
\vspace{-0.3in}
\begin{minted}[fontsize=\footnotesize]{text}
# version QIIME 1
# generate train_rename.fasta from train.fasta
# generate train_taxid.txt from train.fasta
assign_taxonomy.py -i test.fasta -r train_rename.fasta -t train_taxid.txt -m mothur
\end{minted}

\end{spacing}

\begin{figure}[!htbp]
\includegraphics[width=0.98\textwidth]{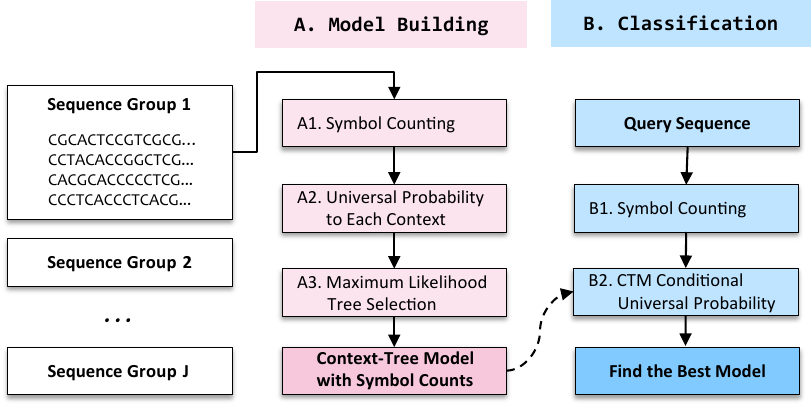}
\caption{\textbf{Overall flow of ~\acl.} The proposed \acl~methodology consists of \textbf{(A)} model-building and \textbf{(B)} classification pipelines that leverage the notion of universal probability in steps~A2 and~B2 in place of the unknown true probability. For each sequence group, the context-tree model with the highest universal probability is found and the query sequence is classified to the group with the highest conditional universal probability given the context tree model.
} \label{fig:overall_flow}
\end{figure}

\clearpage
\section*{\supfigurename~\ref{fig:overall_flow}}
\supspace

\begin{figure}[!htbp]
\includegraphics[width=0.98\textwidth]{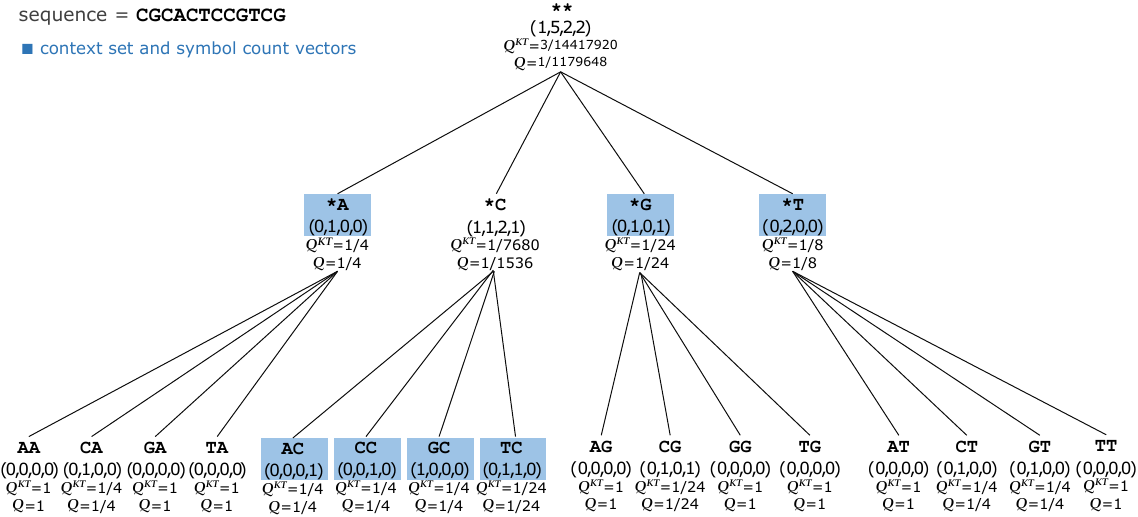}
\caption{\textbf{Context-tree model.}
A complete suffix tree consists contexts from every-order Markov model of $\{\ttS\}^j \times \{\nts\}^{d-j}$ for some $j$, where $\ttS$ denotes ``any''.
Leaf nodes consist of $\{{\nts}\}^d$ and root node consists of $\{{\ttS}\}^d$. Parent nodes have one more $\ttS$ replacing the $\nts$ in child nodes, \eg, $\ttS\ttC$ is a parent node of child-nodes $\ttA\ttC$, $\ttC\ttC$, $\ttG\ttC$, and $\ttT\ttC$. Root node $\ttS\ttS$ is a parent node of child-nodes $\ttS\ttA$, $\ttS\ttC$, $\ttS\ttG$, and $\ttS\ttT$.
A context-tree model (CTM) is a specialized Markov model referred to as a variable-order Markov model. CTM has contexts disjointly, \eg, $\ttS\ttA$,$\ttA\ttC$,$\ttC\ttC$,$\ttG\ttC$,$\ttT\ttC$,$\ttS\ttG$,$\ttS\ttT$.
The $Q^{KT}$ probability is obtained from step A2, and the $Q$ probability is obtained from step A3 in \textbf{\supfigurename~\ref{fig:overall_flow}}.} \label{fig:context_graph}
\end{figure}

\clearpage
\section*{\supfigurename~\ref{fig:context_graph}}
\supspace

\begin{figure}[!htbp]
\includegraphics[width=0.98\textwidth]{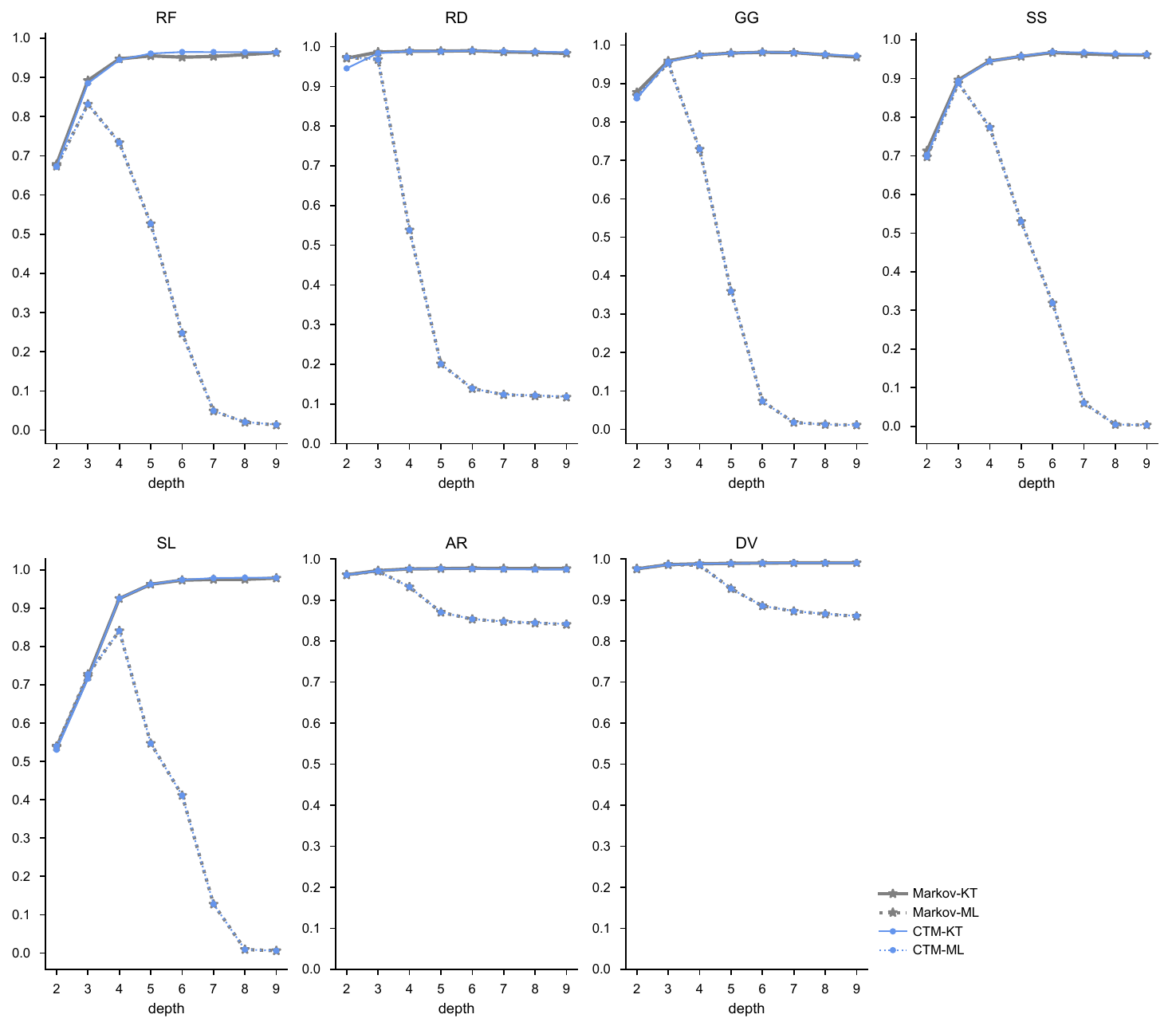}
\caption{} \label{fig:varaccu}
\end{figure}

\clearpage
\section*{\supfigurename~\ref{fig:varaccu}}
\supspace
\vspace{-0.5in}
\noindent\textbf{Accuracy of \acl~variants.}
For each dataset, we examined classification accuracy for the combination of modeling methods
(Markov vs.~context-tree models) and probability estimators (ML vs.~KT). The depth varied from $2$ to $9$.
\acl~ and Markov models are expressed in blue and gray colors, respectively. KT and ML estimators are represented by a solid line and dotted line, respectively. NASCUP (CTM--KT) performed consistently better than other combinations, without any performance degradation when depth becomes too large.

\begin{figure}[!htbp]
\includegraphics[width=0.98\textwidth]{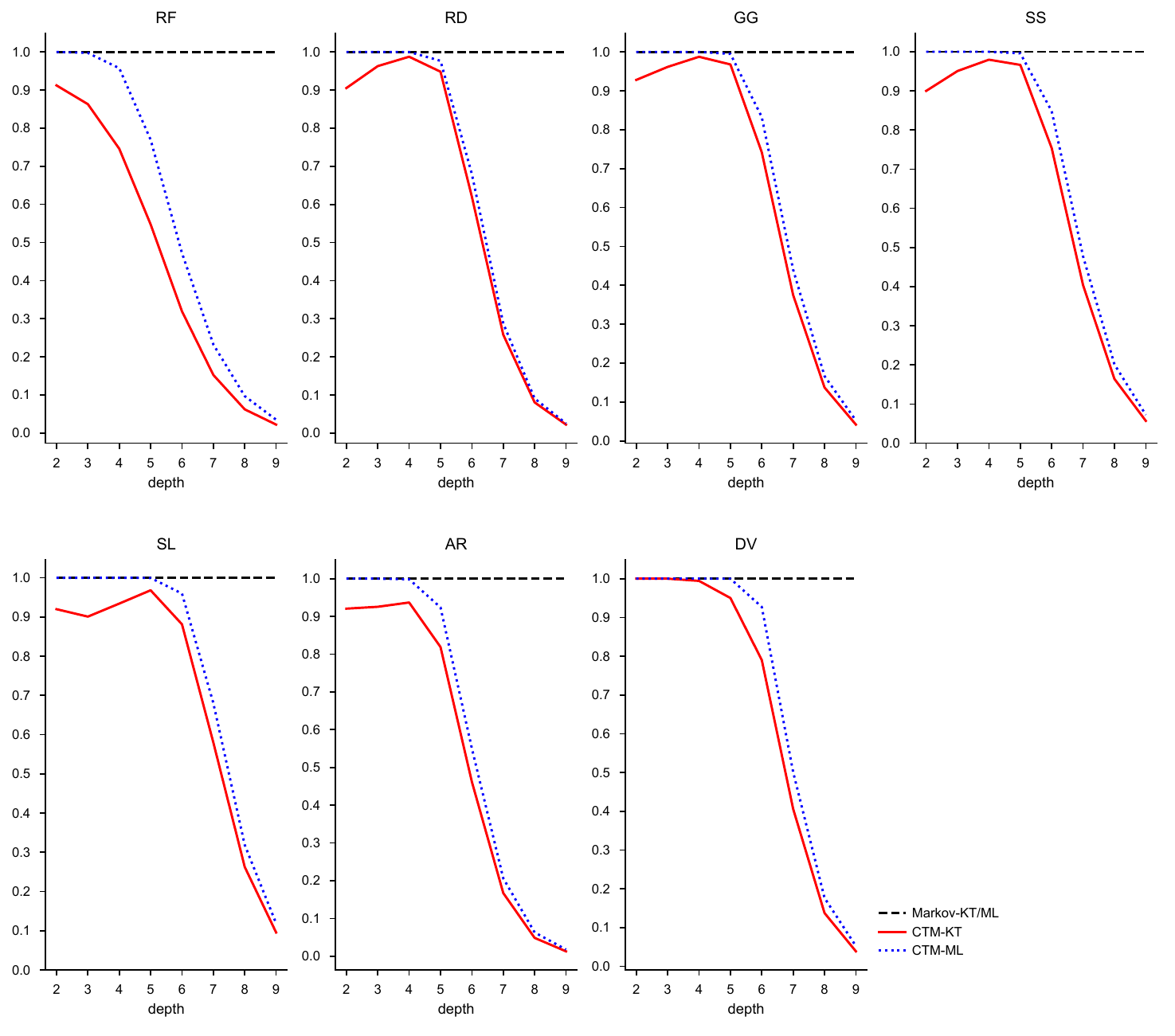}
\caption{} \label{fig:varleaf}
\end{figure}

\clearpage
\section*{\supfigurename~\ref{fig:varleaf}}
\supspace
\vspace{-0.5in}
\noindent\textbf{The ratio of the numbers of leaf nodes.}
The number of leaf nodes in a CTM was much smaller than that of the corresponding Markov model, which demonstrates that CTMs find a sparse and meaningful structure of the sequence groups. For each dataset, the ratio of the average number of leaf nodes of the CTM selected in the model-building stage to the number $4^{d}$ of all leaf nodes in a Markov model is plotted as depth changes. For model building with CTMs, both KT and ML probability estimators were tested, represented by blue dotted and red dashed lines, respectively.

\begin{figure}[!htbp]
\includegraphics[width=0.98\textwidth]{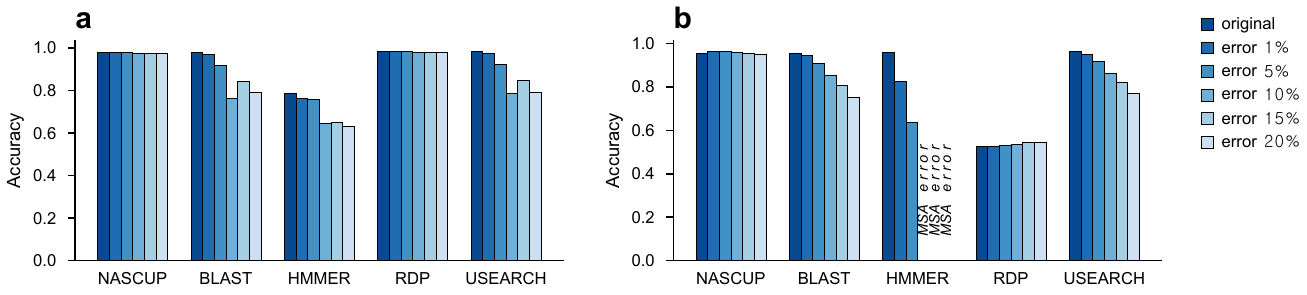}
\caption{} \label{fig:noise}
\end{figure}

\clearpage
\section*{\supfigurename~\ref{fig:noise}}
\supspace
\vspace{-0.5in}
\noindent\textbf{Performance degradation from mislabeling errors.}
Classification accuracy was measured when a fraction of database sequences
in \textbf{(a)} GG and \textbf{(b)} RF datasets
were mislabeled.
The mislabeling error rate increased from 1\% to 20\% and the bar color gradually faded as the error rate increased.
HMMER was not able to build models on RF datasets of higher mislabeling error rate,
because the multiple sequence alignment (MSA) could not be performed
properly due to no consensus columns from too diverse sequences.

\begin{figure}[!htbp]
\begin{subfigure}{1.0\textwidth}
  \includegraphics[width=0.98\textwidth]{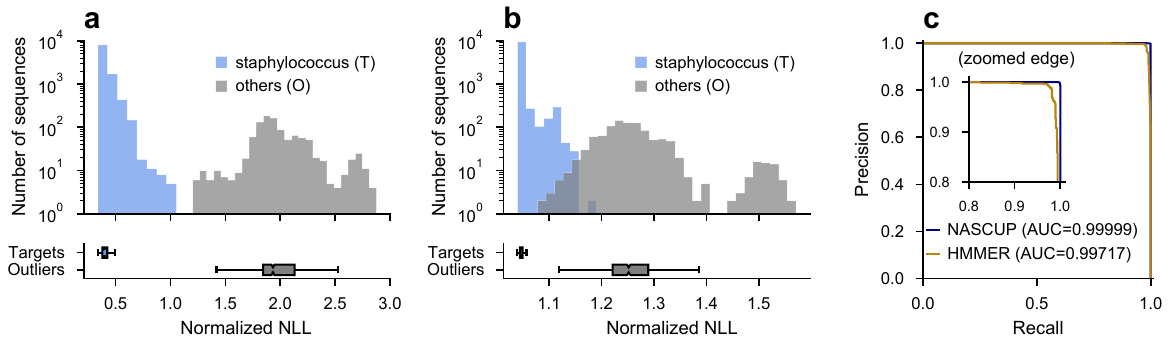}
\end{subfigure}

\hfill

\begin{subfigure}{1.0\textwidth}
  \includegraphics[width=0.98\textwidth]{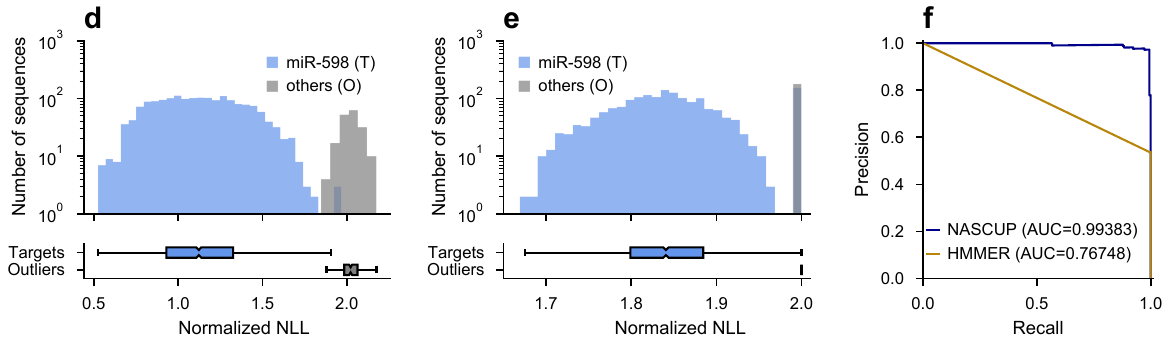}
\end{subfigure}
\caption{}  \label{fig:outlier}
\end{figure}

\clearpage
\section*{\supfigurename~\ref{fig:outlier}}
\supspace

\vspace{-0.5in}
\noindent\textbf{Outlier detection with normalized NLL.}
Distribution of predicted normalized negative log-likelihood (NLL) values with histogram and boxplot of \textbf{(a, d)} \acl~and \textbf{(b, e)} HMMER.
\textbf{(c, f)} Precision-recall (PR) curves and their area under the curve (AUC) values of \acl~and HMMER. The top three plots are experimental results on staphylococcus mixed with outliers from GG dataset, and the bottom three plots are results on miR-598 mixed with outliers from RF dataset.
The datasets were mixed with 10:1 ratio of sequences from targets and others (outliers).
The sequences of outliers were randomly taken from each group except the target group.
The number of sequences in the staphylococcus and others were 10,390 and 1,039, respectively.
The number of sequences in miR-598 and others were 1,800 and 180, respectively.

\begin{figure}[!htbp]
\includegraphics[width=0.98\textwidth]{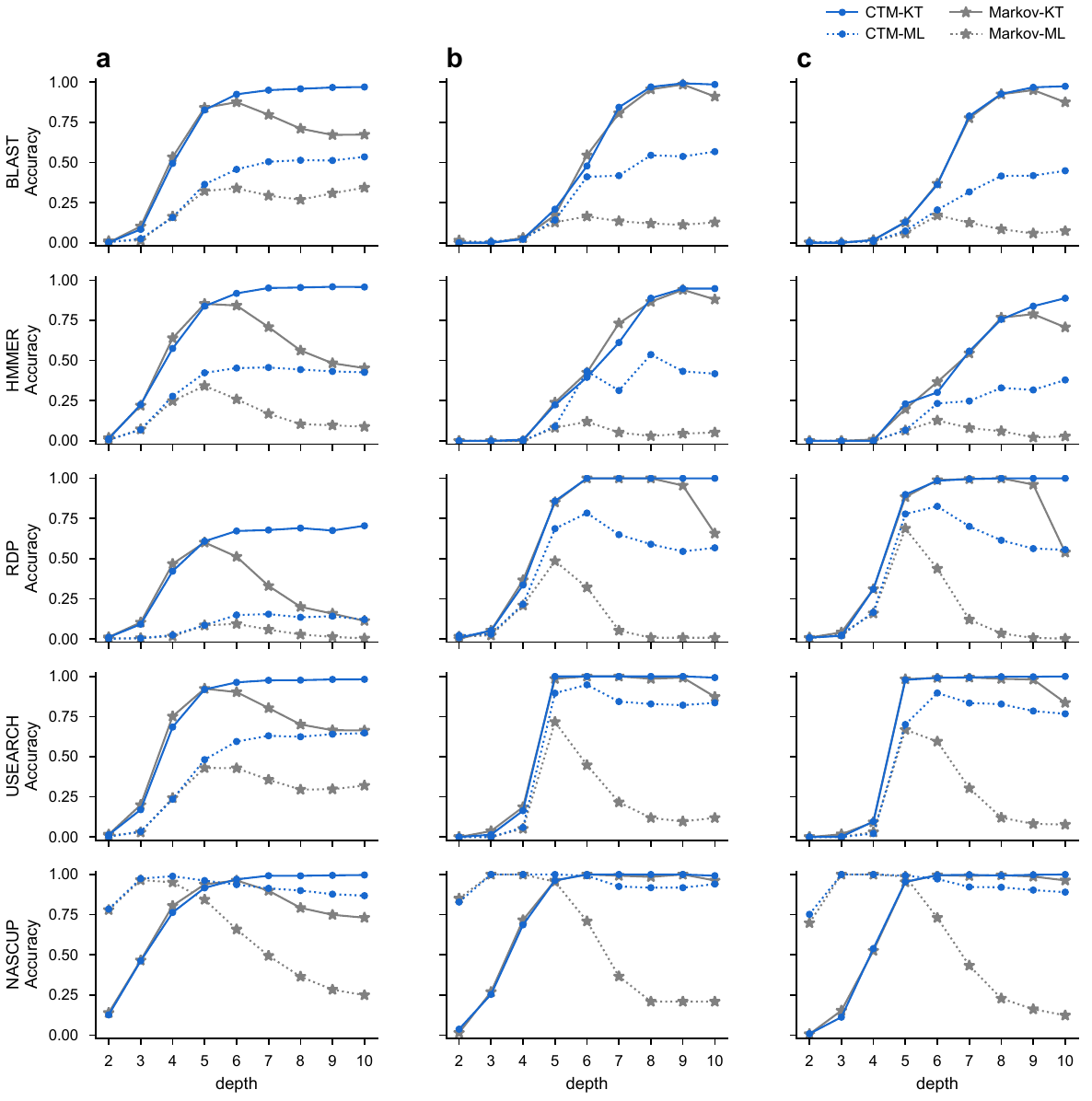}
\caption{}
\label{fig:gen}
\end{figure}

\clearpage
\section*{\supfigurename~\ref{fig:gen}}
\vspace{-0.5in}
\noindent\textbf{Synthetic sequence generation and its quality assessment.}
We generated four types of synthetic sequences by a combination of the modeling methods (Markov and CTM) and sampling methods (universal probability (KT) and maximum likelihood (ML)). With regard to the modeling method, it is represented in a different color (Markov: gray, CTM: blue). With regard to the sampling method, it is represented in a line style (KT sampling: solid line, ML sampling: dotted line). The column represents the classification results according to the datasets of \textbf{(a)} RF, \textbf{(b)} RD, and \textbf{(c)} GG, and the row represents the classification results according to the classification methods in order of BLAST, HMMER, RDP, USEARCH, and \acl. The depth changes from $2$ to $10$.
A total of (a) 1,320, (b) 134, and (c) 464 synthetic sequences (one synthetic sequence per group) were generated and classified.


\clearpage
\section*{\suptablename~\ref{tab:data}}
\textbf{Details of the dataset used in the experiments.} We used a function-based RNA family database named Rfam, taxonomy-based ribosomal RNA databases named RDP, Greengenes, and SILVA, and pyrosequencing data named Artificial and Divergent. The datasets had diverse characteristic: AIGD from 0.08 to 0.33, number of groups from 23 to 1,320, and sequence length from 20 to almost 5,000. The composition of the groups was formed on a completely different purpose by the functional base and the taxonomy base.
\vspace{-0.3in}
\definecolor{LightCyan}{rgb}{1, 0.2, 0.2}

\newcolumntype{C}{>{\centering\arraybackslash}m{15mm}}

\begin{spacing}{\mytablespacing}
\setlength{\tabcolsep}{4pt}

\ctable[
    caption = { },
    label = tab:data,
    pos=htbp,
    star,
    doinside = \footnotesize
]{c@{ }cllcrrr@{--}lc}{
\tnote[$\ast$] {the number of groups with more than 10 preprocessed sequences}
\tnote[$\dagger$]{the total number of sequences after the preprocessing}
\tnote[$\ddag$] {average intra-group distance (the normalized pairwise distance between the sequences within a group), ranging from 0 to 1. The AIGD being close to zero means that the group consists of similar sequences.}
}{
\toprule[1pt]
\multicolumn{2}{c}{classification category} & \multirow{2}[0]{*}{ID} & \multicolumn{1}{c}{dataset name}  & \multirow{2}[0]{*}{AIGD$^\ddag$}  &  \multirow{2}[0]{*}{\# groups$^\ast$\hspace{-1em}} &    \# total$^\dagger\mbox{ }$  & \multicolumn{2}{c}{sequence} &  ground \\
\multicolumn{2}{c}{of the dataset} & & \multicolumn{1}{c}{(version)}   &   &                                          &    sequences   &  \multicolumn{2}{c}{length} &  truth \\

\midrule[1pt]
\multicolumn{2}{c}{functional non-coding RNA} &   RF & Rfam (11.0)~\cite{rfam} & 0.33  &      1,320  & 170,881  &  20&1,875  & accession  \\
\midrule

\multirow{6}{*}{ \begin{turn}{90}{microbial}\end{turn}  \begin{turn}{90}taxonomy\end{turn} }
& & RD &  RDP (10.0)~\cite{methodRDP,schloss2009introducing}   & 0.08  &         134  & 3,838  &  320&1,833  &  \\
&rRNA database & GG & Greengenes (13.5)~\cite{desantis2006greengenes}   &   0.12 &      464  & 23,142  &  1,254&2,146  & taxonomy   \\
&(16S, 18S, 23S/28S) &SS & SILVA-SSU (119.1)~\cite{quast2012silva}   &  0.15  &       313  & 17,625  &  902&3,749  &  (genus level) \\
& & SL & SILVA-LSU (119)~\cite{quast2012silva}   & 0.21   &        107  & 4,593  &  1,900&4,954  &  \\

\cmidrule{2-10}

&pyrosequencing data & AR & Artificial~\cite{quince} & 0.18     &       60  & 44,407  &  40&294 &   reference   \\
&(16S rRNA) & DV &  Divergent~\cite{quince}  &  0.14     &             23  &      55,466  &  38&521  &  sequences  \\


\bottomrule[1pt]
}
\end{spacing}

\definecolor {best}{rgb}{1,0.4,0.7}
\definecolor {poor}{rgb}{0.9,0.9,0.9}
\clearpage
\section*{\suptablename~\ref{tab:result}}
\textbf{Classification accuracy comparison.} Numerical values in the seven dataset columns represent the averages of the correct classification rates under $10$-fold cross validation and the last two columns are the average and geometric means of the averaged values ({\setlength{\fboxsep}{1pt}\colorbox{best}{pink shade}}: best in column; {\setlength{\fboxsep}{1pt}\colorbox{poor}{grey shade}}: unsatisfactory (less than 80\% accuracy)).
\vspace{-0.3in}
\definecolor {highlight}{rgb}{1,0.85,0.85}
\definecolor {best}{rgb}{1,0.4,0.7}
\definecolor {poor}{rgb}{0.9,0.9,0.9}

\begin{spacing}{\mytablespacing}
\setlength{\tabcolsep}{8pt}

\ctable[
    caption = {
    },
    label = tab:result,
    pos=htbp,
    star,
    doinside = \footnotesize
]{r|ccccccc|cc}{
}{

\toprule[1pt]
method/data & RF    & RD    & GG    & SS    & SL    & AR    & DV    & average & geomean \\
\midrule[1pt]
\textbf{\acl} & \cellcolor{best}{\textbf{96.5\%}} & \cellcolor{best}{\textbf{99.0\%}} & 98.2\% & \cellcolor{best}{\textbf{96.9\%}} & 97.4\% & \cellcolor{best}{\textbf{97.6\%}} & 99.0\% & \cellcolor{best}{\textbf{97.8\%}} & \cellcolor{best}{\textbf{97.8\%}} \\
\midrule[0.3pt]
BLAST & 95.8\% & 98.5\% & 98.3\% & 96.3\% & \cellcolor{best}{\textbf{97.4\%}} & 96.5\% & 98.8\%& 97.4\% & 97.4\% \\
USEARCH & 96.5\% & 98.6\% & \cellcolor{best}{\textbf{98.6\%}} & 96.7\% & 97.4\% & 89.8\% & 98.8\%  & 96.6\% & 96.6\% \\
UBLAST &  \cellcolor{poor}{79.9\%} & 98.5\% & 97.9\% & 95.9\% & 97.1\% & 96.4\% & 98.7\%  & 94.9\% & 94.7\% \\
BLAT  &  \cellcolor{poor}{79.1\%} & 97.2\% & 92.1\% & 92.3\% & 95.2\% & 94.7\% & 98.9\%  & 92.8\% & 92.6\% \\
caBLAST & \cellcolor{poor}{39.4\%} &	97.1\%	& 86.9\% &	90.5\% &	93.9\% &	95.5\% & 	97.0\%  &	85.8\% &	82.6\% \\
\midrule[0.3pt]
    UCLUST & \cellcolor{poor}{23.7}\% & 97.0\% & 94.4\% & 85.4\% & 72.6\% & 96.8\% & 98.8\% & 81.3\% & 74.6\% \\
    Mothur & \cellcolor{poor}{52.2}\% & 99.0\% & 98.2\% & 96.3\% & 96.8\% & 95.1\% & 99.0\% & 90.9\% & 89.1\% \\
\midrule[0.3pt]
HMMER & 96.1\% & 98.4\% & 80.1\% & \cellcolor{poor}{14.9\%} & 80.3\% & \cellcolor{poor}{46.3\%} & \cellcolor{poor}{64.0\%}   & 68.6\% & 59.7\% \\
RDP   & \cellcolor{poor}{52.6\%} & 99.0\% & 98.3\% & 96.5\% & 96.9\% & 97.1\% & \cellcolor{best}{\textbf{99.1\%}} & 91.4\% & 89.5\% \\
    QIIME2 & 83.3\% & 98.7\% & 97.1\% & 94.2\% & 95.6\% & 97.0\% & 99.0\% & 95.0\% & 94.8\% \\
Phymm  & 93.6\% & \cellcolor{poor}{77.5\%} & \cellcolor{poor}{76.7\%} & \cellcolor{poor}{39.5\%} & 93.2\% & 95.0\% & 98.9\% & 82.1\% & 79.0\% \\
gzip  & \cellcolor{poor}{62.7\%} & 96.3\% & 90.3\% & 80.1\% & \cellcolor{poor}{77.6\%} & 80.9\% & 96.3\% & 83.5\% & 82.7\% \\
    \bottomrule[1pt]
}

\end{spacing}


\clearpage
\section*{\suptablename~\ref{tab:silva}}
\textbf{Experimental results on large-scale SILVA (SSS, LSU) datasets.}
The sequence groups are formed based on genus-level and species-level from latest SILVA (v132) without compaction.
The genus-level SSU and LSU have 1,861 and 587 groups, respectively, and the species-level SSU and LSU have 3,134 and 749 groups, respectively.
\acl-7 is comparable to RDP whose default $k$ is 8, \acl-6 is the default parameter of \acl.
Each value of accuracy is the average of $10$-fold cross validation.
\vspace{-0.3in}

\begin{spacing}{\mytablespacing}
\setlength{\tabcolsep}{16pt}

\ctable[
    caption = {  },
    label = tab:silva,
    pos=htbp,
    star,
    doinside = \footnotesize
]{rrrrrr}{
\tnote[$\ast$] {\acl-7 means depth of 7 which is comparable to $8$-mer.\\\acl-6 means depth of 6 which is comparable to $7$-mer.}
}{
    \toprule
        \multirow{2}[2]{*}{genus-level}  & \multicolumn{2}{c}{accuracy} & & \multicolumn{2}{c}{time(s)} \\
    \cmidrule(rl){2-3} \cmidrule(rl){5-6}
     & \multicolumn{1}{c}{SSU}    & \multicolumn{1}{c}{LSU}    & &\multicolumn{1}{c}{SSU}    & \multicolumn{1}{c}{LSU} \\
    \midrule
    NASCUP-7$^\ast$ & \textbf{99.04\%} & \textbf{98.68\%} & & 1343.80 & 180.37 \\
    NASCUP-6$^\ast$ & 98.79\% & 98.23\% & &1139.39 & 145.23 \\
    RDP   & 98.76\% & 98.28\% & & 2612.79 & 566.69 \\
    QIIME2 & 97.50\% & 96.73\% & & 2119.97 & 370.37 \\
    UCLUST & 98.05\% & 95.47\% & & 525.62 & 109.92 \\
    Mothur & 98.68\% & 98.23\% & & 9792.79 & 2275.26 \\
    \bottomrule
    \\
    \\
    \toprule
    \multirow{2}[2]{*}{species-level}  & \multicolumn{2}{c}{accuracy} & & \multicolumn{2}{c}{time(s)} \\
    \cmidrule(rl){2-3} \cmidrule(rl){5-6}
     & \multicolumn{1}{c}{SSU}    & \multicolumn{1}{c}{LSU}    & &\multicolumn{1}{c}{SSU}    & \multicolumn{1}{c}{LSU} \\
    \midrule
    NASCUP-7$^\ast$ & \textbf{84.20\%} & 84.78\% &       & 1615.16 & 189.57 \\
    NASCUP-6$^\ast$ & 84.08\% & \textbf{84.88\%} &       & 795.42 & 75.49 \\
    RDP   & 81.25\% & 79.63\% &       & 2605.34 & 323.86 \\
    QIIME2 & 0.41\% & 49.88\% &       & 1937.65 & 229.66 \\
    UCLUST & 78.59\% & 70.48\% &       & 184.91 & 55.16 \\
    Mothur & 79.74\% & 70.97\% &       & 8717.30 & 1279.93 \\
    \bottomrule

}

\end{spacing}

\clearpage
\section*{\suptablename~\ref{tab:time}}
\textbf{Total runtime (classification time) comparison.} Each value in the table including average and geomean is the log-transformed time result (unit: $\log_{10}$ of a second). The classification time for BLAST and its variants is the same as the total runtime, since there is no separate modeling process. For HMMER, the preprocessing time for multiple sequence alignment was excluded. 
\vspace{-0.3in}
\begin{spacing}{\mytablespacing}
\setlength{\tabcolsep}{5.2pt}

\ctable[
    caption = {  },
    label = tab:time,
    pos=htbp,
    star,
    doinside = \footnotesize
]{r|ccccccc|cc}{
}{
\toprule[1pt]
    method/data & RF    & RD    & GG    & SS    & SL    & AR    & DV    & average &geomean \\
\midrule[1pt]      
    NASCUP & 2.5 (2.4) & 1.2 (0.9) & 1.7 (1.5) & 1.5 (1.3) & 1.2 (0.9) & 1.6 (1.0) & 1.6 (1.1) & 1.9 (1.7) &1.6(1.3)\\
\midrule[0.3pt]      
    BLAST & 2.7   & 2.5   & 3.7   & 3.3   & 2.7   & 3.5   & 3.7    & 3.4 &3.2 \\
    USEARCH & 1.2   & 0.2   & 1.4   & 1.1   & 0.6   & 1.2   & 1.3   & 1.1  &1.0\\
    UBLAST & 2.5   & 2.8   & 4.2   & 3.6   & 2.6   & 4.3   & 4.5   & 4.0  & 3.5\\
    BLAT  & 2.6   & 3.5   & 4.8   & 4.5   & 3.6   & 3.3   & 3.3     & 4.2  &3.7\\
    caBLAST & 4.8   & 3.3   & 4.5   & 4.3   & 3.4   & 4.4   & 4.7   & 4.5 & 4.2\\
\midrule[0.3pt]     
 UCLUST & 1.7   & 0.5   & 1.1   & 1.3   & 1.3   & 1.0   & 1.1   & 1.3   & 1.2 \\
 Mothur & 2.5   & 1.2   & 2.1   & 2.0   & 1.6   & 1.4   & 1.5   & 2.0   & 1.8 \\
\midrule[0.3pt]      
    HMMER & 2.9 (2.9) & 3.6 (3.6) & 4.9 (4.9) & 4.8 (4.8) & 4.3 (4.3) & 2.8 (2.8) & 2.5 (2.5) & 4.4 (3.7) &3.7(3.7)\\
    RDP   & 1.6 (1.4) & 0.9 (0.6) & 1.8 (1.5) & 1.6 (1.2) & 1.3 (0.6) & 0.8 (0.4) & 0.8 (0.4) & 1.4 (1.1) &1.3(0.9) \\
QIIME2 & 2.3 (2.0)   & 1.3 (0.8)   & 1.8 (1.2)   & 1.7 (1.1)   & 1.4 (0.9)   & 1.4 (0.9)   & 1.4 (0.9)  & 1.8 (1.4)   & 1.6 (1.1) \\
    Phymm & 3.1 (3.0) & 1.8 (1.5) & 2.8 (2.7) & 2.6 (2.5) & 2.0 (1.7) & 1.7 (1.3) & 1.7 (1.0)  & 2.5 (2.4) &2.2(2.0)\\
    gzip  & 5.2 (5.2) & 3.0 (3.0) & 4.3 (4.3) & 4.1 (4.1) & 3.2 (3.2) & 3.9 (3.9) & 4.0 (4.0) & 4.5 (4.5)&4.0(4.0) \\
\bottomrule[1pt]
}

\end{spacing}

\clearpage
\section*{\suptablename~\ref{tab:data_sc}}
\textbf{Details of the expanding datasets used for scalability experiments.} Expanding datasets were generated from Greengenes\cite{desantis2006greengenes} in two types, sequencewise and groupwise datasets. Sequencewise expansion dataset has fixed number of 1,040 groups and the database size increases gradually. Groupwise expansion dataset has every 104 new groups from the previous phase, and the database size increases gradually as the number of groups and their included sequences increases. Both of the datasets have the fixed number of 5,000 and 1,000 test sequences, respectively.
\vspace{-0.3in}
\begin{spacing}{\mytablespacing}
\setlength{\tabcolsep}{4pt}

\ctable[
    caption = {
},
    label = tab:data_sc,
    pos=htbp,
    star,
    doinside = \footnotesize
]{crrrrrrrrrr}{
\tnote[$\ast$] {the number of training sequences as used database}
\tnote[$\dagger$] {the number of test sequences for classification}
\tnote[$\ddag$] {the number newly added groups or sequences from the previous expanding phase}
}{
\multicolumn{10}{l}{\textbf{Sequencewise expansion dataset}} \\
  \toprule
  Phase & P-1 & P-2 & P-3 & P-4 & P-5 & P-6 & P-7 & P-8 & P-9 & P-10 \\
\midrule
 \# groups &     1,040  &     1,040  &     1,040  &     1,040  &     1,040  &     1,040  &     1,040  &     1,040  &     1,040  &     1,040  \\
           \# tr-seqs$^\ast$ &   55,717  &  111,926  &  161,634  &  204,618  &  279,174  &  336,311  &  391,978  &  434,542  &  495,348  &  555,969  \\
           \# te-seqs$^\dagger$ &     5,000  &     5,000  &     5,000  &     5,000  &     5,000  &     5,000  &     5,000  &     5,000  &     5,000  &     5,000  \\
    \bottomrule
    \multicolumn{10}{l}{ } \\

    \multicolumn{10}{l}{\textbf{Groupwise expansion dataset}} \\
    \toprule
Phase & P-1 & P-2 & P-3 & P-4 & P-5 & P-6 & P-7 & P-8 & P-9 & P-10 \\
\midrule
 \# groups &       104  &       208  &       312  &       416  &       520  &       624  &       728  &       832  &       936  &     1,040  \\
           \# tr-seqs$^\ast$ &   10,842  &   20,339  &   30,818  &   40,739  &   50,457  &   62,287  &   72,602  &   83,847  &   93,984  &  106,003  \\

           \# te-seqs$^\dagger$ &     1,000  &     1,000  &     1,000  &     1,000  &     1,000  &     1,000  &     1,000  &     1,000  &     1,000  &     1,000  \\
    \midrule
\# new$^\ddag$ groups &       104  &       104  &       104  &       104  &       104  &       104  &       104  &       104  &       104  &       104  \\
\# new$^\ddag$ tr-seqs &   10,842  &     9,497  &   10,479  &     9,921  &     9,718  &   11,830  &   10,315  &   11,245  &   10,137  &   12,019  \\

    \bottomrule
}

\end{spacing}

\vspace{-0.1in}

\clearpage
\section*{\suptablename~\ref{tab:bggs_time}}
\textbf{Runtime in the sequencewise expansion datasets.} Since BLAST and USERACH have no modeling procedure, the total time was the same as the classification time. Due to the excessive running time of BLAST and HMMER, the classification time of the two methods was measured proportionally from the experimental results of separate test datasets with only 10 query sequences.
\vspace{-0.3in}
\begin{spacing}{\mytablespacing}
\setlength{\tabcolsep}{8pt}

\ctable[
    caption = {
},
    label = tab:bggs_time,
    pos=htbp,
    star,
    doinside = \footnotesize
]{c|c|rr|rrrrr}{
}{
    \toprule
          & Phase & \# groups & \# tr-seqs & NASCUP & BLAST & HMMER & RDP & USEARCH \\
    \midrule
          & P-1   &      1,040  &        55,717  &            36  &  -    &           759  &             59  &  -  \\
          & P-2   &      1,040  &      111,926  &            65  &  -    &           781  &           107  &  -  \\
          & P-3   &      1,040  &      161,634  &            90  &  -    &           800  &           151  &  -  \\
          & P-4   &      1,040  &      204,618  &          111  &  -    &           819  &           189  &  -  \\
    modeling & P-5   &      1,040  &      279,174  &          149  &  -    &           794  &           256  &  -  \\
    time (s) & P-6   &      1,040  &      336,311  &          179  &  -    &           890  &           306  &  -  \\
     & P-7   &      1,040  &      391,978  &          206  &  -    &           902  &           362  &  -  \\
          & P-8   &      1,040  &      434,542  &          238  &  -    &           909  &           402  &  -  \\
          & P-9   &      1,040  &      495,348  &          291  &  -    &           843  &           453  &  -  \\
          & P-10  &      1,040  &      555,969  &          325  &  -    &           864  &           518  &  -  \\
              \midrule

          & P-1   &      1,040  &        55,717  &            67  &        27,794  &       369,416  &             86  &           132  \\
          & P-2   &      1,040  &      111,926  &            68  &        54,182  &       368,257  &             81  &           394  \\
          & P-3   &      1,040  &      161,634  &            69  &        76,587  &       366,184  &             78  &           507  \\
          & P-4   &      1,040  &      204,618  &            71  &        95,389  &       364,022  &             78  &           630  \\
    classification & P-5   &      1,040  &      279,174  &            70  &       133,472  &       336,251  &             77  &           878  \\
    time (s) & P-6   &      1,040  &      336,311  &            73  &       160,358  &       378,049  &             77  &         1,043  \\
     & P-7   &      1,040  &      391,978  &            74  &       183,837  &       371,104  &             78  &         1,166  \\
          & P-8   &      1,040  &      434,542  &            79  &       203,342  &       367,895  &             82  &         1,285  \\
          & P-9   &      1,040  &      495,348  &            87  &       233,456  &       329,629  &             81  &         1,501  \\
          & P-10  &      1,040  &      555,969  &            88  &       256,936  &       332,026  &             80  &         1,691  \\
                        \midrule

          & P-1   &      1,040  &        55,717  &          104  &        27,794  &       369,416  &           144  &           132  \\
          & P-2   &      1,040  &      111,926  &          132  &        54,182  &       368,257  &           187  &           394  \\
          & P-3   &      1,040  &      161,634  &          159  &        76,587  &       366,184  &           229  &           507  \\
          & P-4   &      1,040  &      204,618  &          182  &        95,389  &       364,022  &           268  &           630  \\
    total & P-5   &      1,040  &      279,174  &          219  &       133,472  &       336,251  &           333  &           878  \\
    time (s) & P-6   &      1,040  &      336,311  &          252  &       160,358  &       378,049  &           383  &         1,043  \\
    & P-7   &      1,040  &      391,978  &          280  &       183,837  &       371,104  &           440  &         1,166  \\
          & P-8   &      1,040  &      434,542  &          317  &       203,342  &       367,895  &           484  &         1,285  \\
          & P-9   &      1,040  &      495,348  &          378  &       233,456  &       329,629  &           535  &         1,501  \\
          & P-10  &      1,040  &      555,969  &          414  &       256,936  &       332,026  &           599  &         1,691  \\
    \bottomrule

}

\end{spacing}

\clearpage
\section*{\suptablename~\ref{tab:bggf_time}}
\textbf{Runtime in the groupwise expansion datasets.} Since BLAST and USERACH have no modeling procedure, the total time was the same as the classification time. Due to the excessive running time of BLAST and HMMER, the classification time of the two methods was measured proportionally from the experimental results of separate test datasets with only 10 query sequences. The model-based methods produce a model for each group independently, so their modeling time was measured based on the newly added 104 groups at each expanding phase.
\vspace{-0.3in}

\begin{spacing}{\mytablespacing}
\setlength{\tabcolsep}{8pt}

\ctable[
    caption = {
},
    label = tab:bggf_time,
    pos=htbp,
    star,
    doinside = \footnotesize
]{c|c|rr|rrrrr}{
}{
    \toprule
          & Phase & \# groups & \# tr-seqs & NASCUP & BLAST & HMMER & RDP & USEARCH \\
    \midrule
          & P-1   &        104  &    10,842  &          12  &  -    &          75  &          12  &  -  \\
          & P-2   &        208  &    20,339  &          12  &  -    &          72  &          11  &  -  \\
          & P-3   &        312  &    30,818  &          12  &  -    &          75  &          12  &  -  \\
          & P-4   &        416  &    40,739  &          12  &  -    &          75  &          11  &  -  \\
    modeling & P-5   &        520  &    50,457  &          12  &  -    &          75  &          11  &  -  \\
    time (s) & P-6   &        624  &    62,287  &          13  &  -    &          76  &          13  &  -  \\
    & P-7   &        728  &    72,602  &          12  &  -    &          75  &          11  &  -  \\
          & P-8   &        832  &    83,847  &          13  &  -    &          75  &          12  &  -  \\
          & P-9   &        936  &    93,984  &          12  &  -    &          75  &          11  &  -  \\
          & P-10  &      1,040  &   106,003  &          14  &  -    &          76  &          13  &  -  \\
              \midrule

          & P-1   &        104  &    10,842  &            9  &       1,406  &       6,718  &            3  &            7  \\
          & P-2   &        208  &    20,339  &          10  &       2,220  &     14,832  &            7  &          14  \\
          & P-3   &        312  &    30,818  &          12  &       3,155  &     24,860  &            8  &          22  \\
          & P-4   &        416  &    40,739  &          13  &       4,006  &     34,355  &          12  &          29  \\
   classification & P-5   &        520  &    50,457  &          15  &       4,922  &     39,643  &          15  &          34  \\
   time (s) & P-6   &        624  &    62,287  &          16  &       5,850  &     46,091  &          20  &          43  \\
   & P-7   &        728  &    72,602  &          18  &       6,762  &     55,132  &          23  &          51  \\
          & P-8   &        832  &    83,847  &          19  &       7,795  &     57,750  &          22  &          59  \\
          & P-9   &        936  &    93,984  &          20  &       8,636  &     74,701  &          25  &          65  \\
          & P-10  &      1,040  &   106,003  &          21  &       9,677  &     76,838  &          30  &          87  \\
              \midrule

          & P-1   &        104  &    10,842  &          21  &       1,406  &       6,793  &          15  &            7  \\
          & P-2   &        208  &    20,339  &          22  &       2,220  &     14,904  &          17  &          14  \\
          & P-3   &        312  &    30,818  &          24  &       3,155  &     24,935  &          19  &          22  \\
          & P-4   &        416  &    40,739  &          25  &       4,006  &     34,430  &          23  &          29  \\
  total & P-5   &        520  &    50,457  &          27  &       4,922  &     39,718  &          26  &          34  \\
 time (s) & P-6   &        624  &    62,287  &          29  &       5,850  &     46,166  &          33  &          43  \\
   & P-7   &        728  &    72,602  &          30  &       6,762  &     55,207  &          34  &          51  \\
          & P-8   &        832  &    83,847  &          32  &       7,795  &     57,824  &          34  &          59  \\
          & P-9   &        936  &    93,984  &          32  &       8,636  &     74,776  &          36  &          65  \\
          & P-10  &      1,040  &   106,003  &          35  &       9,677  &     76,914  &          43  &          87  \\
    \bottomrule
}

\end{spacing}

\clearpage
\section*{\suptablename~\ref{tab:recall}}
\textbf{Accuracy comparison from candidate extraction.} Two sections represent different extracting methods about extracting candidate groups as varying (1) number of candidates, fixed to $K$ and (2) threshold $T$ on the sum of the posterior probabilities, respectively. Each value in the seven datasets is the average accuracy of $10$-fold cross-validation. In the second section, the numbers in parentheses mean the average number of candidate groups.
\vspace{-0.3in}
\begin{spacing}{\mytablespacing}
\setlength{\tabcolsep}{11pt}

\ctable[
    caption = {  },
    label = tab:recall,
    pos=htbp,
    star,
    doinside = \footnotesize
]{r|ccccccc}{
}{
\multicolumn{8}{l}{\textbf{(1) fixed $\boldsymbol{K}$}} \\
\toprule
    $K$ & RF    & RD    & GG    & SS    & SL    & AR    & DV     \\
\midrule
    1  & 0.9648 & 0.9901 & 0.9822 & 0.9685 & 0.9741 & 0.9763 & 0.9902 \\
    3  & 0.9872 & 0.9987 & 0.9986 & 0.9927 & 0.9885 & 0.9867 & 0.9915 \\
    5  & 0.9898 & 0.9990 & 0.9993 & 0.9961 & 0.9915 & 0.9882 & 0.9927 \\
    10 & 0.9926 & 0.9990 & 0.9999 & 0.9977 & 0.9948 & 0.9900 & 0.9942 \\
\bottomrule
\multicolumn{8}{l}{ } \\

\multicolumn{8}{l}{\textbf{(2) threshold $T$ on sum of the posterior probabilities}} \\
\toprule
    $T$ & RF    & RD    & GG    & SS    & SL    & AR    & DV     \\
\midrule
    \multirow{2}{*}{0.95}\quad & 0.9725 & 0.9904 & 0.9828 & 0.9691 & 0.9743 & 0.9823 & 0.9904 \\
	& (1.097) & (1.001) & (1.001) & (1.002) & (1.001) & (1.014) & (1.002) \\
\midrule
    \multirow{2}{*}{0.99}\qquad & 0.9759 & 0.9904 & 0.9832 & 0.9696 & 0.9743 & 0.9840 & 0.9905 \\
    & (1.264) & (1.001) & (1.002) & (1.004) & (1.001) & (1.020) & (1.004) \\
    \midrule
	\multirow{2}{*}{0.999} & 0.9798 & 0.9906 & 0.9838 & 0.9702 & 0.9743 & 0.9854 & 0.9907 \\
	& (1.819) & (1.002) & (1.004) & (1.006) & (1.002) & (1.030) & (1.006) \\
	\midrule
	\multirow{2}{*}{0.9999} & 0.9830 & 0.9906 & 0.9844 & 0.9706 & 0.9747 & 0.9859 & 0.9908 \\
	& (3.004) & (1.002) & (1.006) & (1.008) & (1.003) & (1.053) & (1.008) \\
\bottomrule

}

\end{spacing}

\clearpage
\section*{\suptablename~\ref{tab:stat}}
\textbf{Statistical significance tests.}
For each dataset,
the boldface highlights the p-value of the Wilcoxon signed-rank test for comparing the first and second most accurate classification methods for the dataset.

\vspace{-0.3in}

\begin{spacing}{\mytablespacing}
\setlength{\tabcolsep}{10pt}

\ctable[
    caption = {
},
    label = tab:stat,
    pos=htbp,
    star,
    doinside = \footnotesize
]{rrrrrr}{
}{
    \multicolumn{6}{l}{\textbf{Statistical significance on aggregate datasets}} \\
    \toprule
    \textbf{Normalized} & NASCUP & BLAST & HMMER & RDP   & USEARCH \\
    \midrule
    NASCUP & -     & \textbf{2.0E-03} & 0.0E+00 & 1.61E-295 & 8.0E-16 \\
    BLAST & - & -     & 0.0E+00 & 3.15E-191 & 1.0E-08 \\
    HMMER & - & - & -     & 0.0E+00 & 0.0E+00 \\
    RDP   & - & - & - & -     & 1.52E-117 \\
    USEARCH &- &- & - & - & - \\
    \bottomrule
	\\
	\toprule
    \textbf{Unnormalized} & NASCUP & BLAST & HMMER & RDP   & USEARCH \\
    \midrule
    NASCUP & -     & \textbf{3.9E-11} & 0.0E+00 & 0.0E+00 & 2.3E-31 \\
    BLAST & - & -     & 0.0E+00 & 0.0E+00 & 3.0E-09 \\
    HMMER & - & - & -    & 6.103E-27 & 0.0E+00 \\
    RDP   & - & - & - & -   & 0.0E+00 \\
    USEARCH &- &- & - & - & - \\
    \bottomrule
    \\
    \\
    \multicolumn{6}{l}{\textbf{Statistical significance on individual datasets}} \\
    \toprule
    \textbf{RF}    & NASCUP & BLAST & HMMER & RDP   & USEARCH \\
    \midrule
    NASCUP & -     & 3.7E-06 & 1.8E-02 & 0.0E+00 & \textbf{6.7E-01} \\
    BLAST & - & -     & 1.1E-01 & 0.0E+00 & 1.5E-07 \\
    HMMER & - & - & -     & 0.0E+00 & 4.1E-02 \\
    RDP   & - & - & - & -   & 0.0E+00 \\
    USEARCH &- &- & - & - & - \\
    \bottomrule
    \\
    \toprule
    \textbf{RD}    & NASCUP & BLAST & HMMER & RDP   & USEARCH \\
    \midrule
    NASCUP & -     & 1.8E-01 & 4.6E-02 & \textbf{1.0E+00} & 5.9E-02 \\
    BLAST & - & -     & 7.4E-01 & 1.8E-01 & 3.2E-01 \\
    HMMER & - & - & -     & 4.6E-02 & 7.6E-01 \\
    RDP   & - & - & - & -   & 5.9E-02 \\
    USEARCH &- &- & - & - & - \\
    \bottomrule
}

\ctable[
    caption = {
},
    pos=htbp,
    star,
    doinside = \footnotesize
]{rrrrrr}{
}{
    \toprule
    \textbf{GG}    & NASCUP & BLAST & HMMER & RDP   & USEARCH \\
    \midrule
    NASCUP & -     & 8.9E-01 & 6.6E-92 & 3.3E-02 & 1.3E-01 \\
    BLAST & - & -     & 9.5E-92 & 3.2E-01 & \textbf{8.3E-02} \\
    HMMER & - & - & -     & 9.4E-97 & 1.3E-95 \\
    RDP   & - & - & -  & -     & 7.4E-01 \\
    USEARCH &- &- & - & - & - \\
    \bottomrule
    \\
        \toprule
    \textbf{SS}    & NASCUP & BLAST & HMMER & RDP   & USEARCH \\
    \midrule
    NASCUP & -     & 3.3E-02 & 0.0E+00 & 3.3E-02 & \textbf{4.6E-01} \\
    BLAST & - & -     & 0.0E+00 & 7.2E-01 & 6.3E-02 \\
    HMMER & - & - & -     & 0.0E+00 & 0.0E+00 \\
    RDP   & - & - & -  & -     & 2.8E-01 \\
    USEARCH &- &- & - & - & - \\
    \bottomrule
\\
    \toprule
    \textbf{SL}    & NASCUP & BLAST & HMMER & RDP   & USEARCH \\
    \midrule
    NASCUP & -     & 3.2E-01 & 1.7E-18 & 5.6E-01 & 4.1E-01 \\
    BLAST & - & -     & 2.7E-19 & 1.8E-01 & \textbf{1.0E+00} \\
    HMMER & - & - & -     & 1.3E-17 & 2.7E-19 \\
    RDP   & - & - & -  & -     & 2.6E-01 \\
    USEARCH &- &- & - & - & - \\
    \bottomrule
\\
    \toprule
    \textbf{AR}    & NASCUP & BLAST & HMMER & RDP   & USEARCH \\
    \midrule
    NASCUP & -     & 1.9E-08 & 0.0E+00 & \textbf{4.1E-03} & 4.6E-67 \\
    BLAST & - & -     & 0.0E+00 & 1.0E-04 & 3.3E-47 \\
    HMMER & - & -& -     & 0.0E+00 & 0.0E+00 \\
   RDP   & - & - & -  & -     & 5.3E-60 \\
    USEARCH &- &- & - & - & - \\
    \bottomrule
    \\
    \toprule
    \textbf{DV}    & NASCUP & BLAST & HMMER & RDP   & USEARCH \\
    \midrule
    NASCUP & -     & 2.9E-02 & 0.0E+00 & \textbf{2.1E-01} & 1.6E-06 \\
    BLAST & - & -     & 0.0E+00 & 1.2E-02 & 2.7E-02 \\
    HMMER & - & - & -     & 0.0E+00 & 0.0E+00 \\
    RDP   & - & - & -  & -     & 2.6E-06 \\
    USEARCH &- &- & - & - & - \\
    \bottomrule

}
\end{spacing}

\end{spacing}

\clearpage
\bibliographystyle{naturemag}
\bibliography{reference}